\documentclass[]{aa}

\usepackage{graphicx}
\usepackage{txfonts}
\usepackage[colorlinks=true,citecolor=blue,linkcolor=blue]{hyperref}

\usepackage{natbib}
\bibpunct{(}{)}{;}{a}{}{,} 
\usepackage{bm}
\usepackage{upgreek}
\usepackage[dvipsnames]{xcolor}
\usepackage{amsmath}    
\usepackage{threeparttable}
\usepackage{cancel}
\usepackage{orcidlink}
\usepackage{booktabs}

\newcommand{\lensname}{RXJ1131$-$1231}

\newcommand{\kmps}{km s$^{-1}$}

\newcommand{\as}{$^{\prime \prime}$}

\begin{document} 

    \title{TDCOSMO}
    
    \subtitle{XIX. Measuring stellar velocity dispersion \\ with sub-percent accuracy for cosmography
    }
    
    \titlerunning{Accurate stellar velocity dispersions}
    \authorrunning{S.Knabel, P.Mozumdar et al.}

    \author{
        Shawn Knabel,\inst{1}\thanks{These authors are considered co-first authors.   
        \email{shawnknabel@astro.ucla.edu, pmozumdar@astro.ucla.edu}}
        Pritom Mozumdar\orcidlink{0000-0002-8593-7243},\inst{1}$^{\star}$
        Anowar~J.~Shajib\orcidlink{0000-0002-5558-888X},\inst{2, 3, 4}\fnmsep\thanks{NFHP Einstein Fellow}
            Tommaso~Treu\orcidlink{0000-0002-8460-0390},\inst{1}
            Michele~Cappellari,\inst{5}
            Chiara~Spiniello\inst{5}
            Simon~Birrer\inst{6}
    }
    
    \institute{
    Department of Physics and Astronomy, University of California, Los Angeles, CA 90095, USA; \email{shawnknabel@astro.ucla.edu, pmozumdar@astro.ucla.edu}
    \and
    Department  of  Astronomy  \&  Astrophysics,  University  of Chicago, Chicago, IL 60637, USA; \email{ajshajib@uchicago.edu}
    \and
    Kavli Institute for Cosmological Physics, University of Chicago, Chicago, IL 60637, USA
    \and
    Center for Astronomy, Space Science and Astrophysics, Independent University, Bangladesh, Dhaka 1229, Bangladesh
    \and
    Sub-Department of Astrophysics, Department of Physics, University of Oxford, Denys Wilkinson Building, Keble Road, Oxford, OX1 3RH, UK
    \and 
    Department of Physics and Astronomy, Stony Brook University, Stony Brook, NY 11794, USA
    }
             
   \date{Received xxx, xxxx; accepted xxx, xxxx}

  \abstract
  {The stellar velocity dispersion ($\sigma$) of massive elliptical galaxies is a key ingredient in breaking the mass-sheet degeneracy and obtaining precise and accurate cosmography from gravitational time delays. The relative uncertainty on the Hubble constant H$_0$ is double the relative error on $\sigma$. Therefore, time-delay cosmography imposes much more demanding requirements on the precision and accuracy of $\sigma$ than galaxy studies. While precision can be achieved with an adequate signal-to-noise ratio (S/N), the accuracy  critically depends on key factors such as the elemental abundance and temperature of stellar templates, flux calibration, and wavelength ranges. We carried out a detailed study of the problem using multiple sets of 
  galaxy spectra of massive elliptical galaxies with S/N$\sim$30--160 \AA$^{-1}$, along with state-of-the-art empirical and semi-empirical stellar libraries and stellar population synthesis templates. We show that the choice of stellar library is generally the dominant source of residual systematic errors. We propose a general recipe for mitigating and accounting for residual uncertainties. We show that a sub-percent level of accuracy can be achieved on individual spectra with our data quality, which we subsequently validated with simulated mock datasets. The covariance between velocity dispersions measured for a sample of spectra can also be reduced to sub-percent levels. We recommend this recipe for all applications that require high precision and accurate stellar kinematics. Thus, we have made all the software publicly available to facilitate its implementation.  This recipe will also be used in future TDCOSMO collaboration papers.
}
  

  \maketitle

\section{Introduction}

The stellar velocity dispersion ($\sigma$) has been a key observable for our understanding of massive elliptical galaxies for decades \citep{Faber76}. 
Advances in spectroscopy, stellar populations, and dynamical models have heightened our understanding of their internal structure, orbital anisotropy, mass-to-light ratio, and many other fundamental properties. The tight correlations between stellar velocity dispersion and other parameters serve as a testbed for any galaxy formation model \citep[see a review by][]{Cappellari16}. 

Over the past two decades,  the stellar velocity dispersion has been shown to be particularly powerful in combination with strong gravitational lensing \citep[e.g.,][and references therein]{Treu10b}. 
As a purely dynamical probe, the stellar velocity dispersion allows us to investigate the full three-dimensional structure of galaxies, while remaining insensitive to the mass-sheet degeneracy \citep{Koopmans03}. Conversely, lensing probes the total projected mass \citep{Dutton11} and is insensitive to the mass anisotropy degeneracy \citep{Mamon05,Courteau14}.

For all these reasons, stellar velocity dispersion measurements have assumed a central role in the effort to infer cosmological parameters using gravitational time delays in galaxy-scale lenses; hereafter, this is referred to as time-delay cosmography \citep{Suyu10,Treu16b,Treu22,Birrer24}. By constraining the mass distribution of the deflector (typically a massive elliptical galaxy) \citep{treu02} and by breaking the mass-sheet degeneracy, stellar velocity dispersions can, in principle, help achieve percent-level precision on H$_0$, particularly when spatially resolved kinematics is available \citep{Shajib18, Yildirim20, B+T21}.

However, for velocity dispersions to be a useful ingredient for time-delay cosmography, a level of precision and accuracy better than 1\% is needed, since the relative error on H$_0$ scales as double the error on stellar velocity dispersion, expressed as $\delta H_0 /H_0 = 2 \delta \sigma/ \sigma$ \citep[e.g.][]{Chen21}. Precision is relatively easy to achieve with modern instruments, namely: spectra with current S/N and resolution are more than sufficient to obtain formal errors of order 1\%. Furthermore, if errors are random and uncorrelated, it is sufficient to average multiple systems and/or measurements to obtain the desired precision.  
In contrast, potential systematic errors might bias individual measurements or introduce covariance between different galaxies (or within parts of the same galaxy) thereby introducing an effective systematic noise floor, independent of the S/N or number of systems in the sample. Sources of systematics in stellar velocity dispersion measurements have been a topic of discussion for decades, including instrumental resolution, bias due to insufficient S/N, wavelength range, mismatch between templates and galaxy spectra in terms of elemental abundances and temperature, mismatch between template and galaxy continua, imperfect flux calibration, and interstellar absorption. Many excellent studies on this topic have been published in the past, and their findings inform the work presented here \citep[e.g.,][]{Dressler84,Kelson00,Treu01,Barth02,Spiniello15,Spiniello21a,Spiniello21b,Mehrgan2023,DAGO23}.

In this paper, we revisit the issue of systematic errors in measured stellar velocity dispersion using multiple sets of galaxy spectra with S/N 30-160 \AA$^{-1}$, along with state-of-the-art templates and fitting codes.
Our goal is twofold: minimize systematic errors and quantify residual systematics in a principled, generalized, and reproducible way. Our paper is organized as follows.
In Section~\ref{sec:stars}, we assemble a high-quality set of stellar and stellar population synthesis templates. The high-quality set of templates has been made public for use by third parties. 
In Section~\ref{sec:galaxies}, we present sets of spectra of elliptical galaxies at $z\sim0.1-0.7$ observed with four different instruments \citep[MUSE on VLT, KCWI on Keck, JWST-NIRSpec, SDSS;][]{MUSE, KCWI, NIRSpec}. The SDSS spectra are included for the sake of  comparisons with previous studies. In contrast, the three other sets of spectra are meant to represent  current state-of-the-art instruments used for stellar kinematics of lens galaxies. 

In Section~\ref{sec:systematics}, we examine, quantify, and mitigate all known sources of systematics by running comprehensive sets of measurements.  For all our fits, we use the well-tested and established code pPXF\footnote{\url{https://pypi.org/project/ppxf/}} \citep{Cappellari04,Cappellari17,Cappellari23}. To guard against bugs in the utilization of pPXF and the construction of stellar libraries, SK and PM developed two independent pipelines. Several tests show that the two pipelines yield identical results when applied to the same data.

In Section~\ref{sec:recipe}, we propose a procedure to account for residual systematics by treating unconstrained choices (chiefly the choice of the stellar library) as nuisance parameters and marginalizing over them. The procedure can be used to estimate random and systematic errors as well as covariance between sets of measurements (e.g., spatial bins within a galaxy). 

In Section~\ref{sec:results}, we show that by applying our procedure to high-quality spectra of elliptical galaxies, both systematic errors and covariance between independent galaxies or spatial bins can be contained to a sub-percent level. Our specific results only apply to the conditions described in this paper, in terms of instrumental setup, template libraries, and types of spectra under consideration. However, our procedure is entirely general and can be applied to any set of measurements to estimate the corresponding systematic noise floor. We recommend our recipe for all applications that require high precision and accurate stellar kinematics. We have made all the software publicly available to facilitate its implementation. This recipe will be used in future TDCOSMO collaboration papers. Finally, in Section~\ref{sec:summary}, we summarize and discuss our results.

\section{Stellar libraries and single stellar population models}
\label{sec:stars}

In this section, we briefly describe the stellar libraries used in this work, referring to the original papers for more details (Section \ref{ssec:stars_description}). We then describe our rigorous process to ``clean'' the libraries in Section~\ref{ssec:clean} and compare spectra of the stars in common between the libraries.

\subsection{Description of libraries} \label{ssec:stars_description}

We considered two types of templates for the kinematic extraction: (i) individual stellar spectra and (ii) stellar-population synthesis models based on those stellar libraries. 
As the empirical stellar libraries, we used three that are commonly employed in stellar kinematics determination: Indo-US\footnote{\url{https://noirlab.edu/science/observing-noirlab/observing-kitt-peak/telescope-and-instrument-documentation/cflib}} \citep{Valdes04}, MILES\footnote{\url{https://miles.iac.es/}} \citep{SanchezBlazquez2006,Falcon-Barroso11}, and  X-shooter Spectral Library (XSL)\footnote{\url{http://xsl.u-strasbg.fr/}} \citep{Verro22}. Additionally, we included the sMILES stellar library, a semi-empirical library that was derived by modifying the MILES stellar spectra to represent different $\alpha$ element abundances \citep{Knowles21}. This modification is meant to address template mismatch issues often encountered in the study of massive elliptical galaxies \citep{Barth02, Mehrgan2023}. However, it is important to note that these models have known limitations, particularly in the blue part of the optical spectrum \citep{Knowles21, Knowles23}.  Therefore, our expectation is that they would not be sufficiently accurate for 1\% kinematics. We included them in our study for completeness.

For our stellar population synthesis models, also known as single stellar population (SSP) models, we used the E-MILES and sSMILES, (hereafter E-MILES SSP and sMILES SSP, \citealt{Vazdekis10, Vazdekis16, Knowles23}) as well as the SSP model based on the XSL library (hereafter XSL SSP, \citealt{Verro22b}). While these models offer a physically motivated mix of stars, they lack the flexibility provided by individual stellar spectra. Therefore, they cannot reproduce the complexity of real star formation histories and chemical enrichment histories and are unlikely to be sufficient for sub-percent accuracy. However, they have been popular choices in the past. Therefore, we investigated them for completeness to determine how well they agree with the stellar libraries. 

The Indo-US and MILES libraries have an approximately constant full width at half maximum (FWHM) kinematic resolution of 1.36 \AA\ and 2.5 \AA, respectively \citep{Falcon-Barroso11}, which correspond to $\sigma_{\text{t}} \approx$ 43 km s$^{-1}$ and 80 km s$^{-1}$, respectively, at 4000 \AA. For all MILES-based libraries (E-MILES, sMILES, etc.), we adopted 2.5 \AA\ as the FWHM resolution of the templates. The kinematic resolution of the XSL templates is $\sigma_{\rm t} \approx 13$ km s$^{-1}$ at the UVB region (3000--5560 \AA), 11 km s$^{-1}$ at the VIS region (5330--10200 \AA), and 16 km s$^{-1}$ at the NIR region (9940--24800 \AA). We refer to the original papers for a full description of the libraries.

\subsection{Cleaning the stellar libraries}
\label{ssec:clean}

The full stellar libraries contain spectra of variable quality, including some with defects, incomplete data, incomplete calibration, or low S/N. Indeed, when using stellar spectra to construct stellar population synthesis models, the original papers applied quality cuts \citep{Vazdekis16, Verro22}. 

To obtain ``clean'' spectral libraries, we took a series of steps. First, we eliminated all the spectra that were flagged by the original teams as faulty. For example, for XSL, the template FITS file headers contain keywords that indicate the applied corrections and quality flags and some of the spectra were flagged as ``peculiar,'' ``abnormal,'' and ``zero'' \citep[see, e.g., Tables B.1 and C.1 of][]{Verro22}. In addition, some spectra have not been corrected for slit losses. We eliminated from XSL all spectra listed in the aforementioned Table C.1 and all that were corrected for slit loss. Whenever possible, we contacted the authors of the library to reproduce the selection applied to their construction of the stellar population synthesis models.

Second, we eliminated all the spectra with incomplete stellar parameters, especially the temperature, T$_{\rm eff}$, and metallicity, Fe/H. For MILES and XSL, they represent a small fraction (1-2\%) of the total number of stars, while for Indo-US, they are a larger fraction (20\%). In almost all cases, they are faulty spectra (e.g., missing part of the spectral range, very noisy, or with obvious defects). We note that many of the spectra flagged by the authors of the MILES and XSL libraries are also missing $T_{\rm eff}$ and Fe/H values. These two initial steps remove between 20-30\% of the initial spectra for all three libraries. For reference, the Indo-US library has been used in the most recent versions of the SDSS kinematic pipeline \citep[][and Bolton, A.S. private communication]{Bolton08}. However, to our knowledge, no cleaning was applied.

\begin{figure}
\includegraphics[width=\columnwidth]{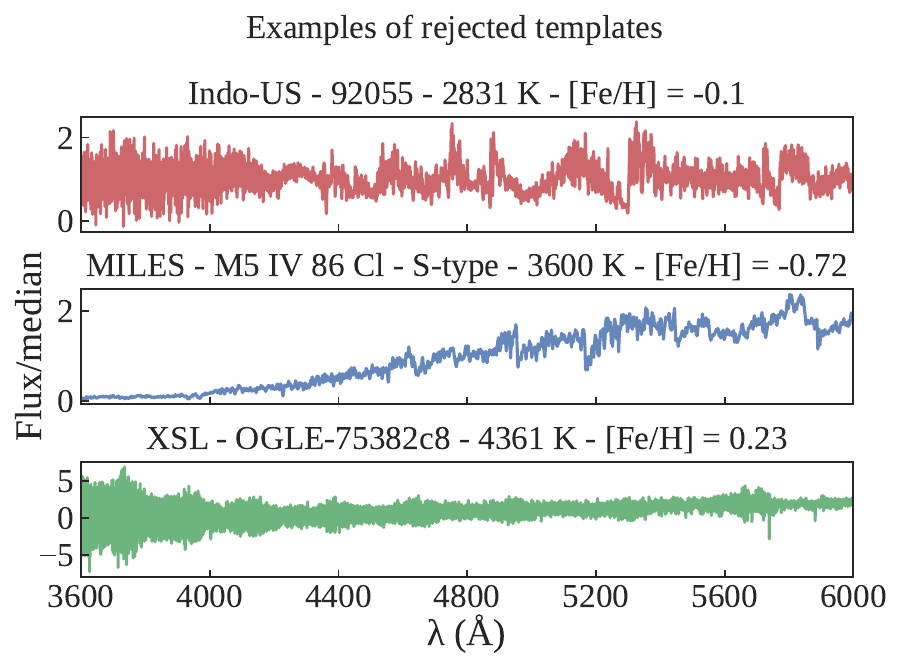}
        \caption{\label{fig:rejected_templates}
        Example of rejected templates from each of the three stellar libraries. For each spectrum, we state the library name, object name, $ T_{\rm eff}$, and metallicity above the respective panel. All three templates were removed because they were too noisy in the wavelength region of interest.} 
\end{figure}

Third, we visually inspected all the individual spectra to identify defects or anomalies. Although the general quality of the spectra is excellent, in a few cases, substantial anomalies were found (see Table~\ref{table:cleaning} and Figure~\ref{fig:rejected_templates}). The spectra were flagged for the following anomalies:

\begin{enumerate}
\item Improper flux calibration at the edges and near the dichroic. These spectra were kept in the clean libraries. However, we recommend avoiding or masking the regions near the edge and near the dichroic when performing the fits.
\item Improper correction of telluric absorption. These spectra were kept in the clean libraries. However, we recommend masking out the regions corresponding to the telluric A and B bands for high-accuracy work.
\item Some spectra (most commonly those coming from the Indo-US library) have localized gaps that can extend for 100s or 1000s of \AA. We did not exclude them from the clean library, but we recommend excluding them if the gaps overlap with the fitted range. 
\item Some spectra show clear emission lines, which could be due to the contamination by other astronomical targets along the line of sight. Those have been removed from the clean library, as they modify the stellar absorption lines with infilling.
\item Some spectra are much noisier than the galaxy spectra we consider in this work (see examples in Figure~\ref{fig:rejected_templates}). They have been removed from the clean library as they are too noisy for precision work.
\item A very small number of spectra have clear defects, possibly resulting from imperfect reduction or calibration (e.g., there is one spectrum for which the flux of the extinction corrected spectrum is below that of the original one in the red). Those have been excluded from the clean library.
\end{enumerate}

\begin{figure*}
\includegraphics[width=\textwidth]{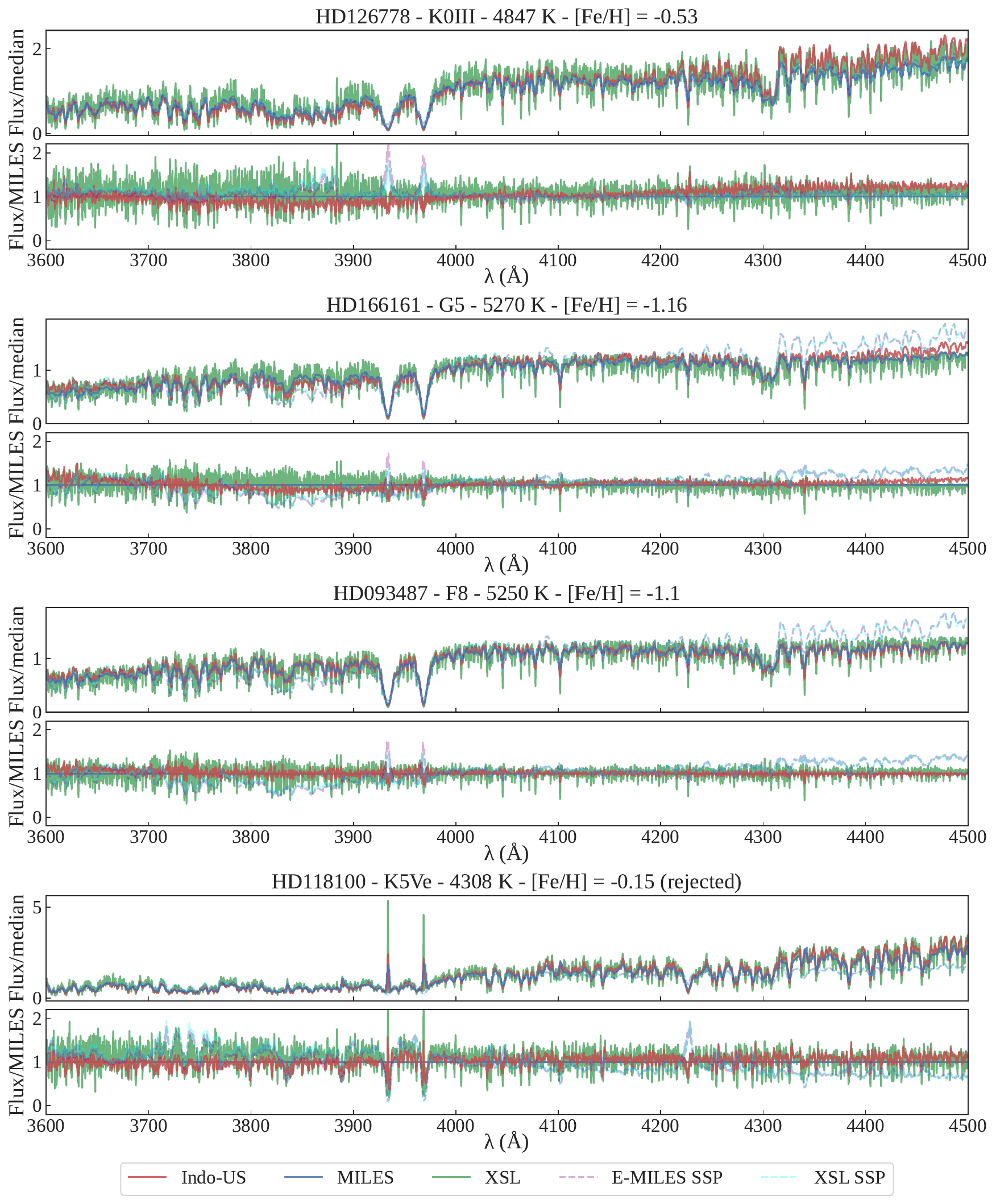}
        \caption{\label{fig:stellar-spectra}
        Example spectra of stars in common between the three libraries. For each example, the spectra are shown in the top panel of the two, and ratios with respect to the MILES spectrum are shown in the bottom one. SSP spectra with solar metallicity and an age of 10 Gyr are also shown for comparison. The fourth template shown (HD118100) was removed from all three clean libraries, owing to the strong emission lines in the CaHK absorption features. It was removed from MILES by author selection and was flagged independently by visual inspection in the other two libraries.}
\end{figure*}

After the cleaning procedure, the stellar libraries consist of 789, 496, and 992 stars, as summarized in Table~\ref{table:cleaning}. \footnote{The list of stars in the ``clean'' stellar libraries has been made available in electronic format at \url{https://github.com/TDCOSMO/KINEMATICS_METHODS}.}
For the clean stars in common between the libraries, we compared the spectra visually and in terms of their ratio with respect to the MILES spectrum for each star. Some examples are shown in Figure~\ref{fig:stellar-spectra}. Overall, the comparison is excellent, even though there are some instances in which the ratio differs significantly from one, indicating systematic residuals in the flux calibration and reddening corrections. These differences underscore the importance of having additive and multiplicative polynomial continuum components in the kinematic fits, as discussed below.

\begin{table}
\caption{Number of stars in stellar libraries at each step of cleaning. }
\begin{center}
\begin{threeparttable}
\begin{tabular}{l l l l}
    \hline
Step    & \# MILES &  \# XSL  & \# Indo-US \\
\hline
Initial &    985      & 830         &    1273        \\
After library authors' cuts\tablefootmark{a} &  801   & 530         &     1273       \\
With $T_{\rm eff}$ and Fe/H\tablefootmark{b} & 790 &  522       &    1006        \\  
After a visual inspection & \textbf{789} & \textbf{496}    &   \textbf{992}         \\  
\hline
\end{tabular}
\end{threeparttable}
\end{center}
\label{table:cleaning}
\tablefoot{
\tablefoottext{a}{Quality cuts applied by the authors of the libraries in the construction of SSP models \citep{Vazdekis16, Verro22}}
\tablefoottext{b}{Stars that do not have consistent estimates of $T_{\rm eff}$ and Fe/H (usually poor-quality spectra)}
}
\end{table}

\subsection{Robustness to template distribution biases} \label{ssec:template_robustness}

\begin{figure*}
\centering
\includegraphics[height=50mm]{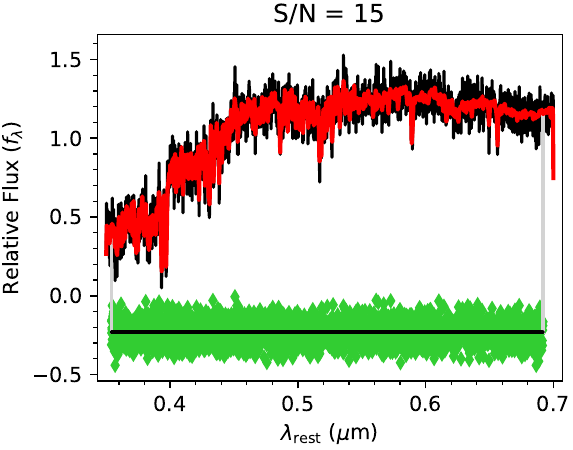}
\includegraphics[height=47mm]{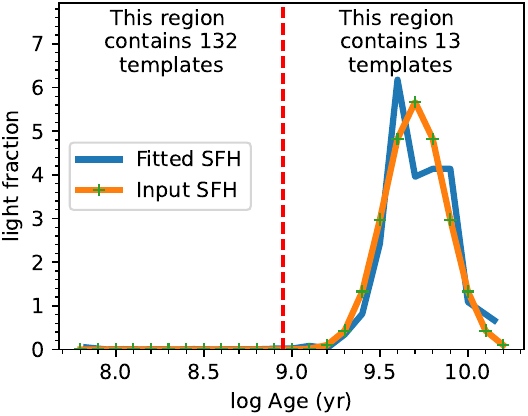}
\includegraphics[height=47mm,trim=27 0 0 0,clip]{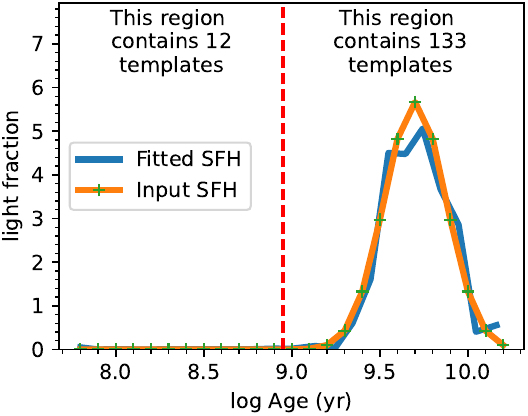}

    \caption{\label{fig:Test-templates}
    Test of the robustness of the spectral fit to the template distribution, as described in Section~\ref{ssec:template_robustness}. The left panel shows a representative pPXF fit to the mock spectrum, showing the data (black), best-fit model (red), and residuals (green diamonds). The fits for the two biased library cases are indistinguishable, except for different random noise realizations. The middle and right panels show the input SFH (orange) and the averaged recovered pPXF SFH (blue) of 100 Monte Carlo realizations for the cases with template libraries overpopulated at young and old ages, respectively. The recovered SFHs are statistically indistinguishable, demonstrating the robustness of the method.
        }
\end{figure*}

To obtain an unbiased estimate of the stellar velocity dispersion, it is crucial to use stellar libraries that span the expected temperature and metallicity ranges. We first verified that our quality-control cuts on the stellar libraries do not alter the parameter distribution, ensuring our sample remains representative.

A potential concern is that the distribution of template parameters within a library could bias the spectral fit, by acting as a prior on the recovered parameters. In this section, we show that pPXF is not affected by this issue. To test the robustness of our method, we designed an extreme experiment in which we intentionally overpopulated specific regions of the parameter space, that is, the population parameters of the spectral library.

We first constructed a mock spectrum by normalizing all E-MILES SSP templates to a unit mean flux, then co-adding different templates with weights that depend on the template age. The weights follow a Gaussian distribution centered at $\lg\text{Age}=9.7$ with $\lg\sigma_\text{Age}=0.2$, where Age is givwn in years, and we adopted an intrinsic dispersion of 100 km/s.

For clarity, we restricted the templates to solar metallicity, using the E-MILES library of stellar population models \citep{Vazdekis16} and created a logarithmically spaced grid of stellar ages. We then generated two biased template libraries by artificially increasing the density of templates by a factor of 10, either for ages younger or older than the logarithmic midpoint (defined as $\lg\mathrm{Age} \geq 9$, with Age in years), using linear interpolation in $\lg\mathrm{Age}$. Each of these biased libraries was then used with pPXF to fit the mock spectrum, using four additive polynomials and without any regularization. For each set of templates, we performed 100 random realizations of the noise to average out the noise in the recovered star formation history (SFH).

As shown in Figure~\ref{fig:Test-templates}, the averaged recovered star formation histories were statistically indistinguishable for both biased libraries. Although this experiment does not directly test the kinematics, the insensitivity of the recovered star formation history to these extreme variations provides a strong demonstration of the method’s robustness. This result implies that as long as the relevant parameter space (i.e., the population of the spectral library) is adequately covered, the fit will consistently recover the same best-fitting template and, by extension, the same kinematics, regardless of the input template distribution.

\section{Galaxy samples} 
\label{sec:galaxies}
 
Our goal is to investigate systematic errors in the measurement of stellar kinematics of massive elliptical galaxies, which comprise the vast majority of deflectors in galaxy-scale strong gravitational lenses. To be sensitive to percent level effects in the systematic errors and covariance, we need spectra with sufficiently high S/N from galaxies that act as gravitational lenses -- or that reside within the appropriate redshift and luminosity ranges that typically characterize strong gravitational lenses. 

Our first sample is a set of massive elliptical galaxies observed with the MUSE spectrograph \citep{Bacon2010} on VLT, described in Section~\ref{ssec:MUSE}. Our second sample is a set of gravitational lens galaxies observed with KCWI on the Keck Telescope, described in Section~\ref{ssec:KCWI}. We also consider the high-quality spectrum of the lens galaxy \lensname\ observed with JWST-NIRSpec \citep{NIRSpec}, as described in Section~\ref{ssec:JWST}. Even though it is only one galaxy, there are many spectral bins within the galaxy, and thus, we can observe the covariance between the spectra. Furthermore, NIRSpec covers the rest frame CaT region, 
which is difficult to reach from the ground at our redshifts of interest.
Thus, NIRSpec allows us to quantify systematics in a different wavelength region than that of MUSE and KCWI. 
We also included a sample of spectra taken by the Sloan collaboration (Section~\ref{ssec:SDSS}). Those generally have a much lower S/N than the other spectra considered here. However, they have been broadly used in the literature. Therefore, we believe it is valuable to demonstrate how our considerations might apply to the SDSS dataset.

Throughout this section, we used a S/N metric to compare the quality of data across different instruments. Our chosen metric is the average S/N per \AA, measured in the rest frame wavelength interval of 4000--4500 \AA\ (chosen to be redward of the 4000 \AA\ break, where the S/N per pixel is relatively uniform). This is obtained by averaging the S/N per pixel, divided by the square root of the number of \AA\ per pixel. This metric is not applicable to our JWST data, owing to the spectral range covered. For that dataset, we modify the wavelength range to 8303-8838~\AA.
When making comparisons across different wavelength ranges and surveys, it is also interesting to consider the S/N per 70 \kmps\ interval (the SDSS pixel resolution). At our fiducial pivot point 4250 \AA, 70\ \kmps\ is almost exactly 1 \AA. 

\subsection{VLT-MUSE}
\label{ssec:MUSE}

\begin{figure*}

\sidecaption
    \includegraphics[width=12cm]{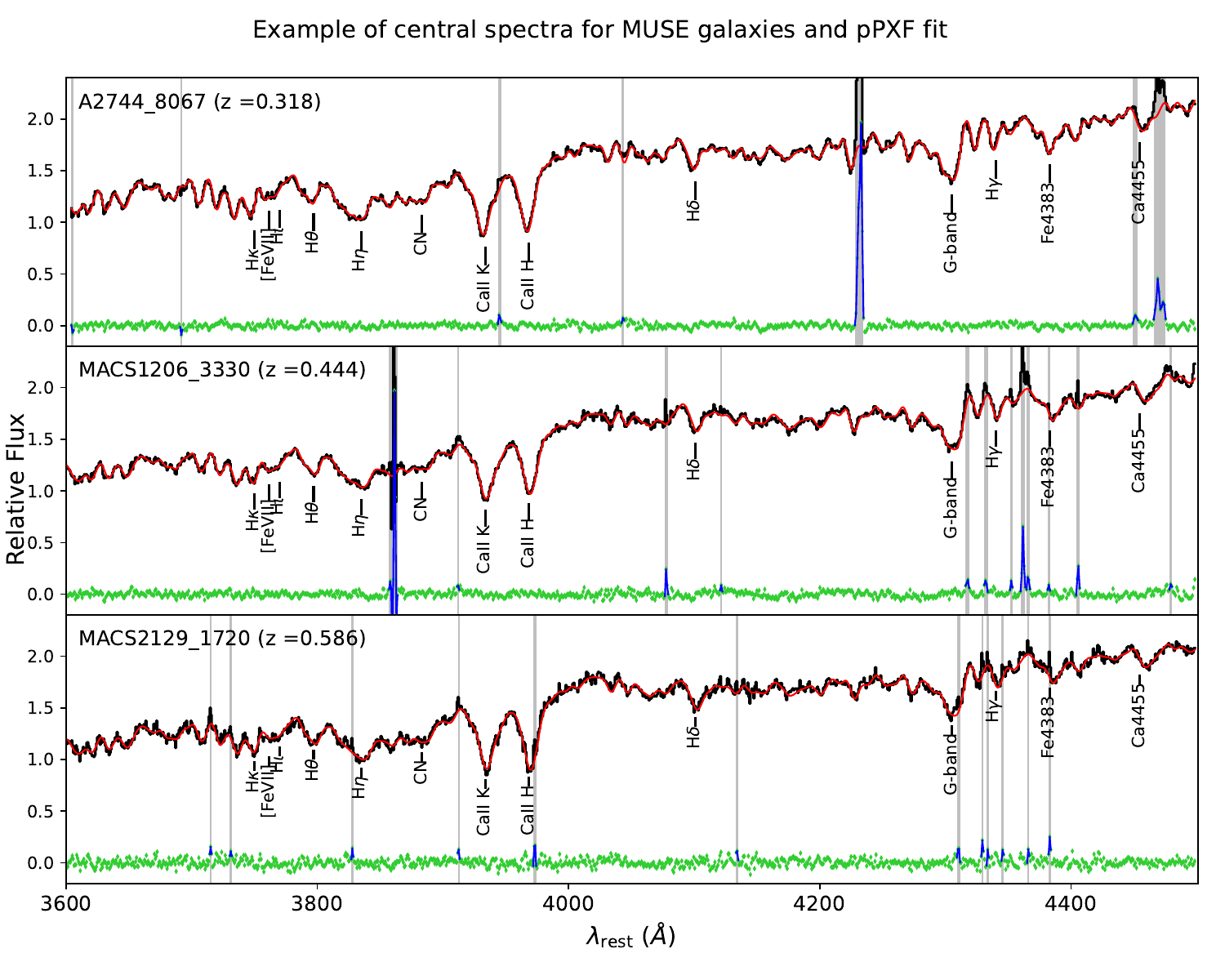}
    \caption{\label{fig:MUSE_galaxy_spectra}
        1D spectra extracted from central spaxels of three galaxies from the MUSE sample at different redshifts. In each panel, the black line is the observed spectrum and the red line is the best-fit model from pPXF. The gray regions mark the excluded wavelength range from fitting, and the green markers are residuals between data and the fitted model at each wavelength. The blue pixels, generally emission lines or sky residuals, have been excluded by sigma-clipping.}
\end{figure*}

The Multi-Unit Spectroscopic Explorer (MUSE) is an integral-field spectrograph situated at the Very Large Telescope (VLT) of the European Southern Observatory (ESO). It provides a spatial sampling of 0.2\as across a 1$^{\prime 2}$ field of view (FoV) in its wide-field mode \citep{Bacon2010}. It operates within a wavelength range of $\lambda =$ 4650-9300~\AA\ with a reciprocal dispersion of 1.25~\AA\ per pixel. The FWHM spectral resolution is approximately 2.5 \AA\ around 5000 \AA\ which corresponds to an instrumental dispersion of 65 km $\mathrm{s}^{-1}$. The galaxies in the original MUSE sample are selected using the integral field unit (IFU) datacubes from the ESO Science Archive under the MUSE-DEEP program \footnote{\url{https://doi.org/10.18727/archive/42}}, which consolidates observations from single or multiple MUSE programs at specific sky locations into a single IFU datacube, achieving very high effective exposure times. The selection criteria, data extraction process, and other relevant information about the sample, such as the galaxy ID and the galaxy cluster or field to which it belongs, ESO archive ID of the datacube, program IDs of the observations used to create the datacube, effective exposure, and other details are described in \citet[][]{MAGNUS_I}. From the parent sample of \citet{MAGNUS_I}, we first selected a subsample of 140 galaxies that contain data within our adopted restframe wavelength range of 3600-4500~\AA\ with S/N$>$30 \AA$^{-1}$.
The central spectra of these galaxies were visually inspected to ensure that no reduction artifacts were present. A handful of spectra were removed due to reduction artifacts, poor sky subtraction, and the presence of strong emission lines.  
Finally, a velocity dispersion cut of $\sigma_{v} >$ 140 km s$^{-1}$ was applied to the sample. The reason for this cut is described in Section~\ref{ssec:templates_resolution}. In total, the final sample consists of 85 galaxies with an average integrated S/N of 60 \AA$^{-1}$. The redshift range of this sample is 0.29-0.67. Some examples of the central spectra are shown in~Figure~\ref{fig:MUSE_galaxy_spectra}.

\subsection{Keck-KCWI}
\label{ssec:KCWI}

\begin{figure*}
\sidecaption
    \includegraphics[width=12cm]{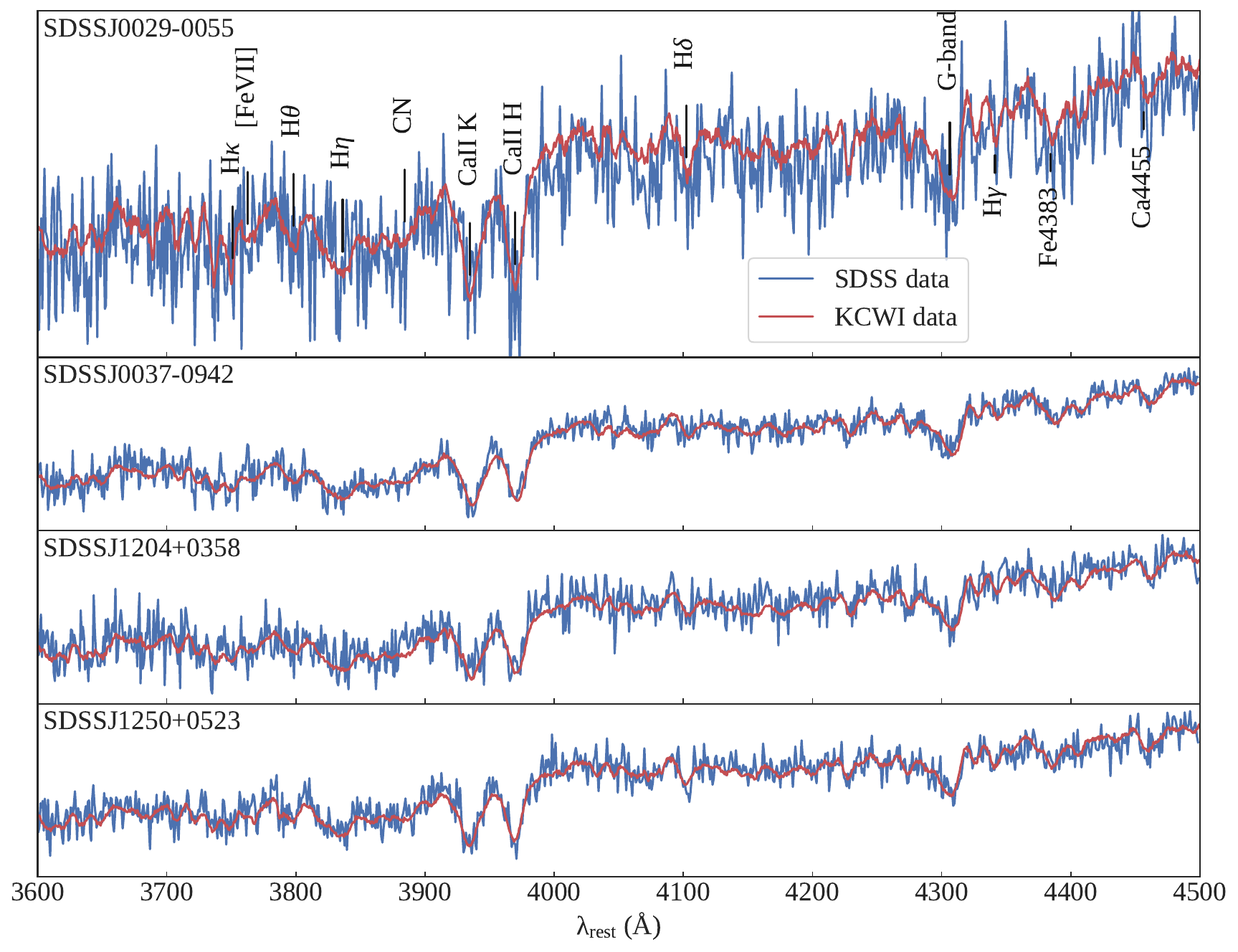}
    \caption{Comparison between the KCWI (red) and SDSS (blue) spectra of the same objects in the wavelength range used for the KCWI fit. The KCWI spectra shown here are integrated over spaxels within a radius of $1.5\arcsec$, the same size as the SDSS fiber. SDSS spectra (blue) shown here have S/N $<15\ \text{\AA}^{-1}$.}\label{fig:KCWI_SDSS_galaxy_spectra}
\end{figure*}

Fourteen of the Sloan Lens ACS Survey (SLACS) gravitational lenses \citep{Bolton06,Bolton08} were observed with KCWI on Keck, with the goal of measuring spatially resolved stellar kinematics. The data are fully described by \citet{Knabel24}. The data are collected in the wavelength range 3500-5600~\AA\ in a $8.4''\times20.4''$ FoV, and the spectral resolution of R $\sim$ 3600 corresponds to an instrumental dispersion $\sigma_{\rm inst} = 35$ km $\mathrm{s}^{-1}$. The reciprocal dispersion is 0.5~\AA\ per pixel. The redshift range of the sample is 0.15-0.35. In order to attain a high S/N and avoid contamination from the lensed background source, we integrated the datacube within a circular radius of $1''$ centered on the center of the foreground deflector galaxy. For most of our studies (and unless otherwise specified), we used these $1''$-integrated spectra. The average S/N of the sample is 160 \AA$^{-1}$ within this radius. For comparison with SDSS spectra of these objects, we also extracted spectra from the KCWI datacube integrated within the SDSS aperture-radius of $1.5''$. These spectra are shown in Figure~\ref{fig:KCWI_SDSS_galaxy_spectra} together with the corresponding SDSS spectra for comparison. We use these spectra exclusively for comparison with SDSS. We note that the SDSS spectra for this subsample of objects have insufficient S/Ns for our purposes and were therefore not used in the overall analysis. We did, however, use a set of higher S/N spectra from SDSS, as described below.

\subsection{JWST/NIRSpec}
\label{ssec:JWST}

\begin{figure*}
\includegraphics[width=\textwidth]{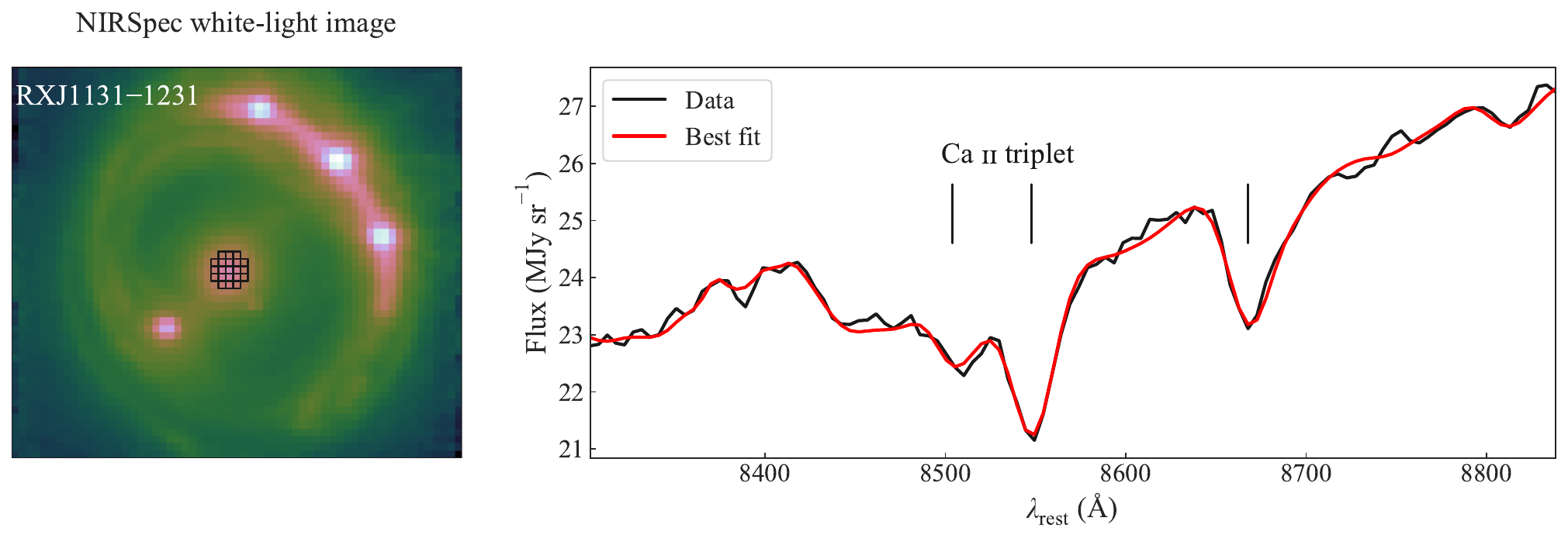}
        \caption{\label{fig:nirspec_spectra}
        JWST/NIRSpec observations of the lens galaxy RXJ1131$-$1231. \textbf{Left:} White-light image from the NIRSpec datacube. The 21 spaxels at the center of the lens galaxy, over which we average our test results, are traced with black squares. \textbf{Right}:  Spectrum (black) from the central spaxel and the best fit (red) from \textsc{pPXF} using the ``clean'' Indo-US library. The Ca \textsc{ii} triplets are marked with vertical lines.}
\end{figure*}

The gravitational lens \lensname\ was observed by JWST/NIRSpec in IFS mode \citep{NIRSpec22} during Cycle 1 (Program 1794; PI Suyu, Co-PI Treu). The data analysis and processing are fully described by \citet{Shajib25}. We constructed the reduced datacube with a spaxel size of 0\farcs1. The redshift of the lens is $z=0.295$. The white-light image of the NIRSpec datacube and one example spectrum is shown in Figure~\ref{fig:nirspec_spectra}. 

We took 21 single spaxels within the central $0\farcs5\times0\farcs5$ region, excluding the four at the corners, of the lens galaxy (illustrated in Figure~\ref{fig:nirspec_spectra}), over which we average our test results. We restricted the bins within the central region of the lens galaxy to minimize contamination from the background quasar and its host galaxy in the spectra adopted for the test. Additionally, we selected a wavelength range (8303--8838 \AA) sufficiently encompassing the Ca \textsc{ii} triplet lines. This region also includes the H$\alpha$ broad emission + [N \textsc{ii}] lines from the quasar plus host galaxy appearing slightly blueward of the Ca \textsc{ii} triplet and the [S \textsc{ii}] lines from the host in between the two prominent Ca \textsc{ii} lines at $\sim$8600~\AA. We modeled all these lines from the quasar and its host with additional kinematic components in our fit \citep[for details, see][]{Shajib25}. However, for the tests performed in this paper, we only focused on the extracted kinematics from the lens galaxy's Ca \textsc{ii} triplet features.

We removed some spectra from the Indo-US and XSL stellar template libraries that have \texttt{NaN} values or coverage gaps within the fitted wavelength range, leaving us with 609 and 489 templates in them, respectively. We did not include the MILES stellar template library for the tests performed on the NIRSpec data as it does not cover the Ca \textsc{ii} triplet's wavelength region. 

When extracting kinematics from the IFU data, it is standard practice in the literature \citep[e.g.,][]{Manga19,Shajib23} to first synthesize a single global template. This single global template is obtained from the best-fit weighted combination of the whole template library when fitting to the spectra summed from multiple central spaxels or bins, which has a considerably larger S/N. Afterward, all the individual bins are fit using this single global template. This approach reduces bin-to-bin scatter in the extracted kinematic map for a single galaxy. The creation of a global template from a central region of the galaxy is included as an additional step to the methodology described in the previous section. We find that this additional step indeed reduces bin-to-bin scatter and, thus, in turn, the systematic scatter between the template libraries. Although this is helpful in reducing systematic scatter for the adopted case of NIRSpec spectra, this step might not be necessary for cases with a a much higher S/N per spatial bin, where contamination from quasar or host galaxy lines can be robustly differentiated, unlike the specific case presented in this paper. 

\subsection{SDSS}
\label{ssec:SDSS}

\begin{figure*}
\sidecaption
    \includegraphics[width=12cm]{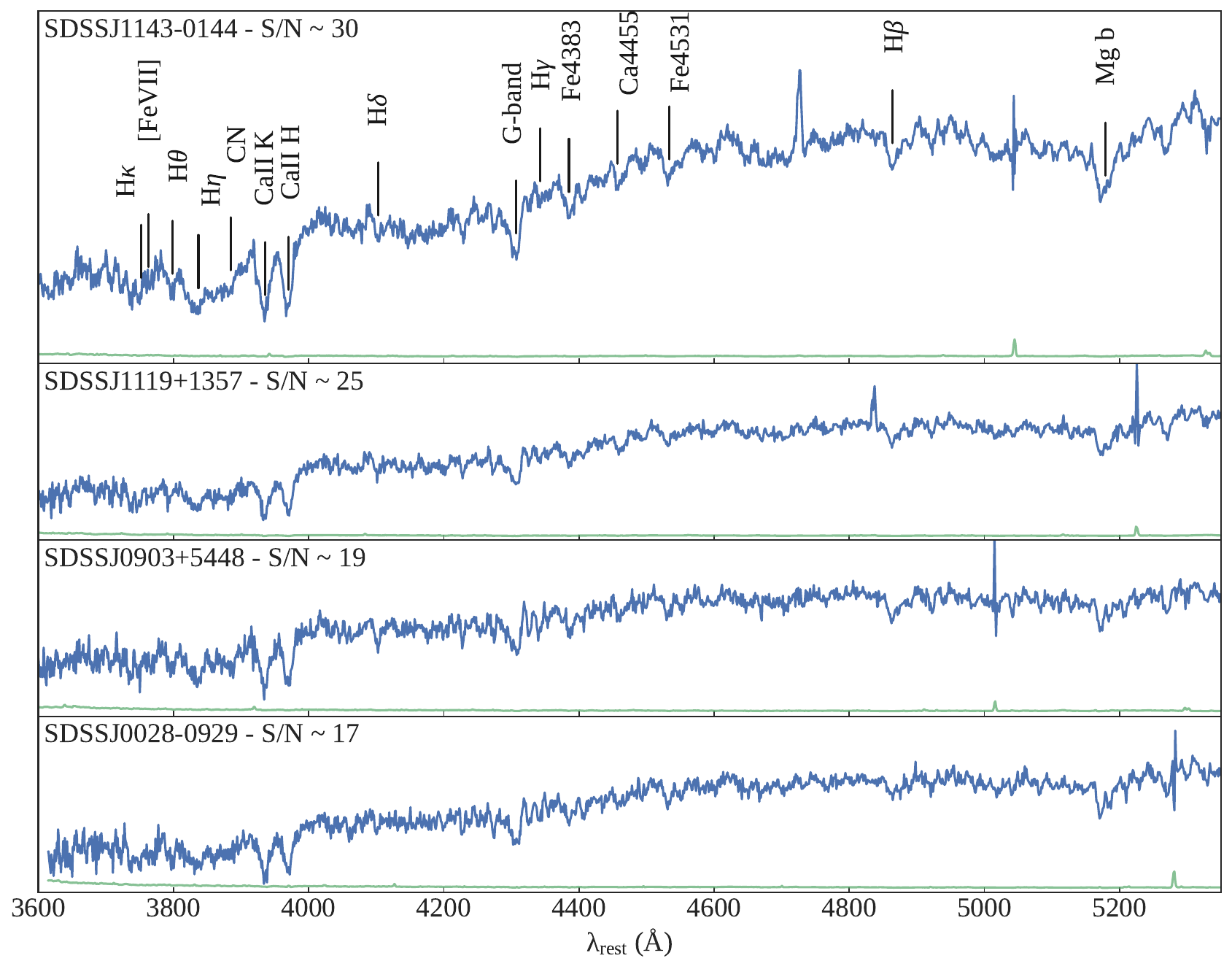}
    \caption{Examples of SDSS spectra with S/N$>15\ \text{\AA}^{-1}$, which is needed to measure kinematic extraction at the level of precision and accuracy we require in this study. The wavelength range shown is the range for fits involving the SDSS data. We take this range instead of using the entire SDSS spectrum because fits with XSL templates that cross the UVB-VIS dichroic above 5450 \AA\ are unreliable. \label{fig:SDSS_spectra}}
\end{figure*}

For reference and comparison, we also consider SDSS spectra of a sample of galaxies that were identified by \cite{Bolton08, Shu15} as potential gravitational lenses. Targets were then selected from this list for follow-up with the Hubble Space Telescope, and the confirmed lenses represent the SLACS \citep{Bolton08} and ``SLACS for the masses'' \citep{Shu15} lenses. Multiple velocity dispersion measurements are available for this sample from the early SDSS releases, the SLACS papers themselves, and the most recent data release, DR17 \citep{DR17}. We refer to \citet{Knabel24} for a detailed comparison between the KCWI and SDSS spectra. In general, the SDSS spectra have much lower S/N than required for the present analysis, as shown in Figure~\ref{fig:KCWI_SDSS_galaxy_spectra}. However, 50 of them have S/N $>15$ \AA$^{-1}$. We present results for those here.  An example of this set of relatively high-S/N SDSS spectra is shown in Figure~\ref{fig:SDSS_spectra}. This subsample of 50 SLACS lens candidates spans a lower redshift range than the KCWI sample ($\rm z \sim 0.04-0.15$) with an average S/N $\sim19$ \AA$^{-1}$.

\section{Testing for systematics}
\label{sec:systematics}

In this section, we describe a number of systematic tests conducted on our galaxy samples. As we  explain here, the dominant source of systematic errors turns out to be the choice of stellar templates, provided that reasonable choices are made for wavelength range and continuum polynomials. We first discuss systematic errors associated with the spectral resolution of the templates in Section~\ref{ssec:templates_resolution}. Then we discuss continuum polynomials (Section~\ref{ssec:polynomials}), cleaning templates (Section~\ref{ssec:clean_test}), choice of templates (Section~\ref{ssec:templates}), and wavelength range (Section~\ref{ssec:wave}). 

\subsection{Spectral resolution of the templates}
\label{ssec:templates_resolution}

In practice, the stellar velocity dispersion $\sigma$ is inferred by comparing the data to a template. If the template resolution is matched to the data, then the astrophysical $\sigma$ is the one returned by \textsc{ppxf} without the need for changes. Templates of higher resolution than the data are initially broadened to match the spectral resolution of the instrumental setup, to find the additional broadening imposed by the stellar kinematics.  For templates with lower resolution than the observed data, the effect of this broadening must be applied as a correction after measuring the dispersion from the data.

Under a Gaussian approximation, the instrumental resolution of the spectrograph used to observe the template $\sigma_{\rm t}$, and that used to observe the galaxy spectrum $\sigma_{\rm s}$, are related to the total smoothing of the observed spectrum $\sigma_{\rm o}$ and the astrophysical stellar velocity dispersion $\sigma$ via the equation 

\begin{equation}
\sigma^2=\sigma_{\rm o}^2 - \sigma_s^2 - \sigma_t^2
\label{eq:sigma_quad}
.\end{equation}

For small errors in the characterization of $\sigma_{\rm s}$ and $\sigma_{\rm t}$, it is easy to show -- by expansion to leading order -- that the relative variations of the astrophysical velocity dispersion and the instrumental and template resolution measurements are related by the following equations. The relative variation in the astrophysical dispersion, at fixed $\sigma_{\rm o}$ and $\sigma_{\rm s}$ is,

\begin{equation}
\frac{\delta \sigma}{\sigma}\approx \frac{\delta \sigma_{\rm t}}{\sigma_{\rm t}}\left(\frac{\sigma_{\rm t}}{\sigma}\right)^2,
\label{eq:sigmat}
\end{equation}

While the relative variation at fixed $\sigma_{\rm o}$ and $\sigma_{\rm t}$ is
\begin{equation}
\frac{\delta \sigma}{\sigma}\approx \frac{\delta \sigma_{\rm s}}{\sigma_{\rm s}}\left(\frac{\sigma_{\rm s}}{\sigma}\right)^2.
\label{eq:sigmas}
\end{equation}

These variations are generally sub-percent. For example, even for a 10\% error in the determination of an instrumental resolution of 50 km s$^{-1}$, the corresponding error on a velocity dispersion of 250 km s$^{-1}$ is 0.4\%. If they are random (e.g., stemming from instabilities of the spectral configuration of slit-filling effects in variable seeing), they will average out for a sample. If they are not random, one should ensure that they are below the noise floor required by the measurement. 

For example, the MILES library has a resolution of approximately $\sigma_t \sim 80$ \kmps at 4000 \AA; thus, if the goal is 1\% errors on $\sigma$, and assuming that the error on $\sigma_t$ is 3\% \citep{Falcon-Barroso11}, and according to Eq,~\ref{eq:sigmat}, it should not be used for stellar velocity dispersions below 140 \kmps.  In contrast, if the goal is 5\%, MILES can be used down to 60 \kmps. Since the goal of this paper is 1\% we discarded galaxies for which the higher resolution libraries (e.g. XSL) give velocity dispersions below 140 \kmps, because MILES would not meet our standards of precision for this paper and therefore we would not be able to carry out the comparison between libraries.

\subsection{Continuum and flux calibration polynomials}
\label{ssec:polynomials}

The fitting code pPXF uses additive and multiplicative polynomials to account for mismatches between the templates and the galaxy spectra. Additive polynomials can represent light from non-stellar sources (e.g., an active nucleus), imperfect sky subtraction, or variations in continuum line strength to reduce template mismatch. Multiplicative polynomials can account for improper flux calibration due to slit losses or differential atmospheric dispersion. 

Additive polynomials are less computationally expensive than multiplicative ones and are often preferred for large datasets. However, the computational cost of multiplicative polynomials is still manageable for the sample described here, taking just a few minutes per galaxy. Our extensive tests show that there is usually a ``sweet spot'' in polynomial orders, where the results from the fit are very stable with respect to specific choices of templates or wavelength ranges. Choosing significantly lower or higher orders can result in significant discrepancies between the results. Therefore, we recommend running a grid of polynomial orders to identify the most appropriate ones for a specific dataset, depending on the wavelength range and accuracy of the spectrophotometric calibration \citep[see, e.g.,][for a similar strategy]{DAGO23}.

In practice, we find that additive polynomials of order 6 (as well as 5 or 7) and multiplicative polynomials of order of 2 (as well as 1 or 3) are appropriate for the adopted wavelength range (described in Section \ref{ssec:wave}) for our MUSE/KCWI/SDSS datasets. For JWST-NIRSpec, we adopted an additive polynomial order of $n=1$ and a multiplicative polynomial order of $m=0$, owing to the shorter wavelength range and flatter overall continuum.  As we show in the next section, the residual systematic uncertainty associated with polynomial order choice is at the level of 0.2-0.5\% for the datasets considered here.

\subsection{Cleaning the libraries}
\label{ssec:clean_test}

\begin{figure*}
\sidecaption
    \includegraphics[width=12cm]{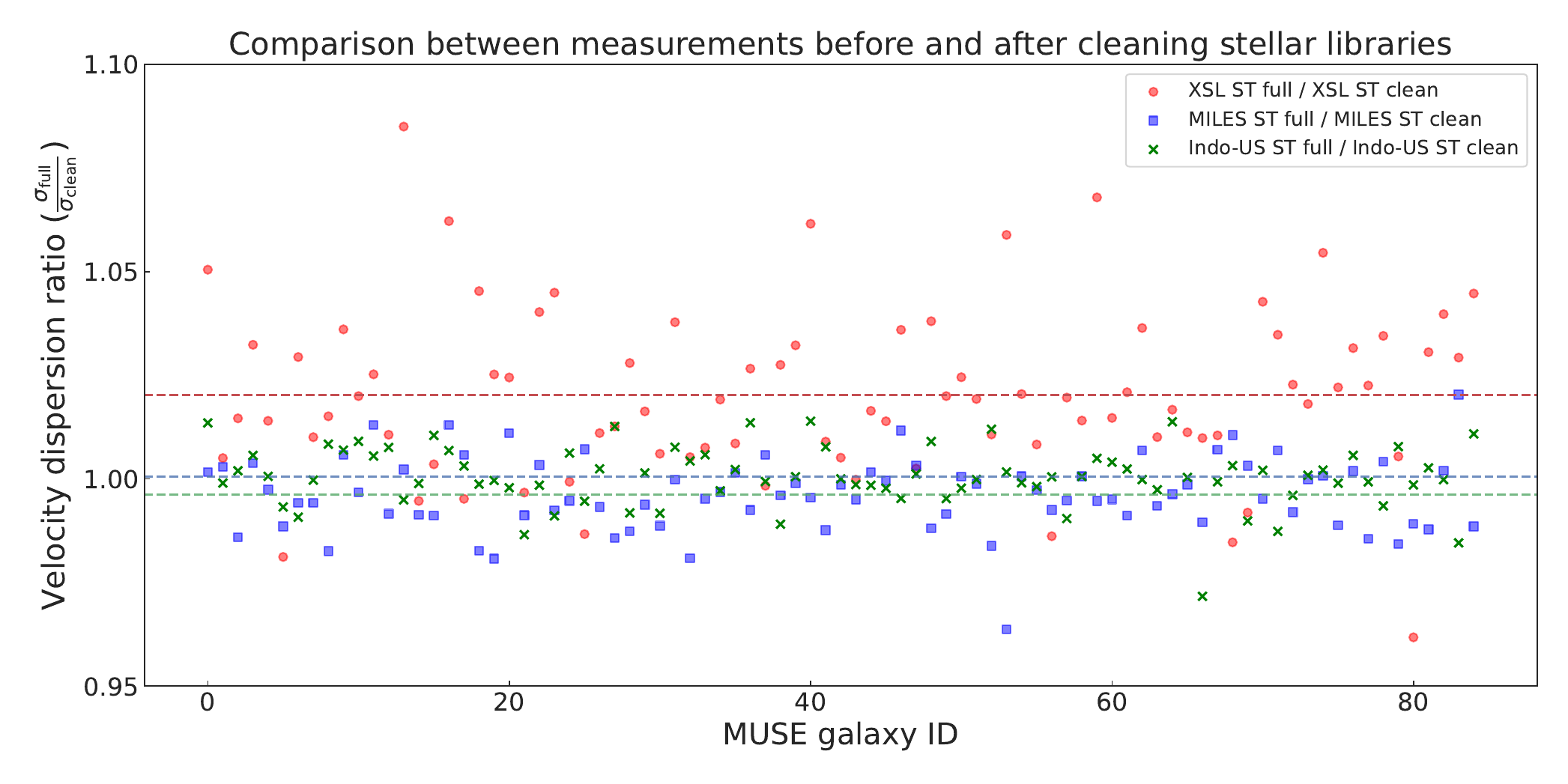}
        \caption{\label{fig:comparison_full_vs_clean_library}
        Ratio of velocity dispersion measurements between "full" and "clean" XSL (red), MILES (blue), and Indo-US (green) libraries. The markers denote the ratio of the velocity dispersion measurements for individual galaxies in the MUSE sample and the horizontal dashed lines mark the sample average for each library. While the effect of cleaning the libraries is most prominent for XSL, it is evident that this is a crucial process in achieving high accuracy.}
\end{figure*}

The effect of cleaning the libraries is significant. Flawed template spectra can cause a variety of issues with the resulting model fit. For example, in the stellar spectrum in the lowest panel of Figure~\ref{fig:stellar-spectra}, the emission lines in the centers of the CaHK absorption features could artificially broaden and reshape the line profiles in combination with other stellar templates used in the fit. In Figure~\ref{fig:comparison_full_vs_clean_library}, we show the ratio of velocity dispersion measurements of the MUSE sample using `full'\footnote{We note that there are some spectra in the full libraries which contain only 0 and \texttt{NaN}. Those need to be excluded to prevent pPXF from crashing.} and `clean' XSL, MILES, and Indo-US libraries (the results for the other samples are similar). 
We see that the effect is of order percent on average for MILES and Indo-US, but it is significantly larger for XSL. In some cases, the differences can be several percent, up to 5\% or more. It is thus crucial to use the cleaned libraries for any high-precision measurement.

\subsubsection{Prior on stellar temperature and metallicity} 

\begin{figure*}
\includegraphics[width=\textwidth]{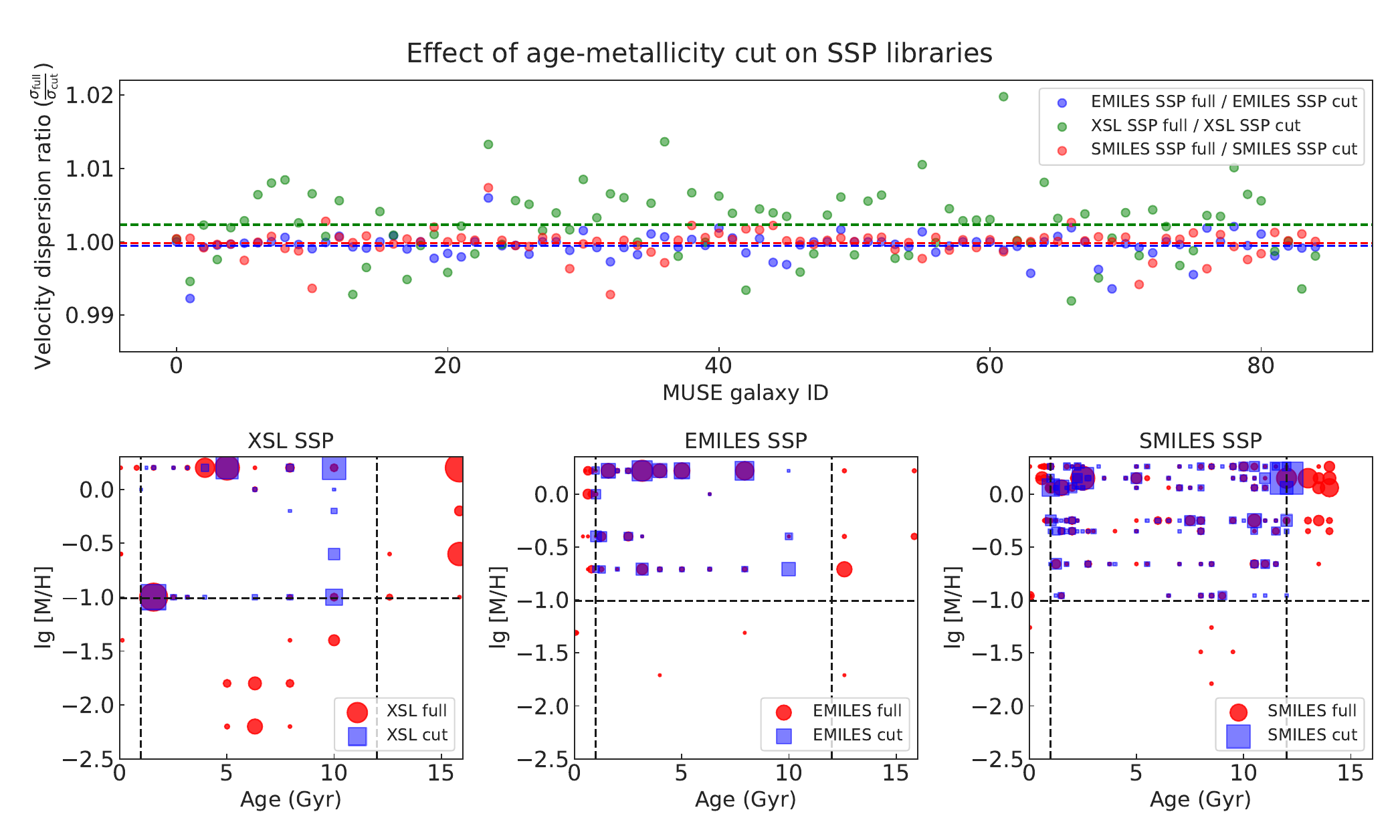}
        \caption{\label{fig:effect_age_metallicity_cut_SSP}
        Comparison of velocity dispersion measurements of the MUSE sample using full E-MILES, sMILES, and XSL SSP libraries vs.~the same libraries after applying cuts on age and metallicity of the templates. The top panel shows the ratio of velocity dispersion measurements using the `full' and `cut' SSP libraries. The blue (E-MILES), green (XSL), and red (sMILES) markers show the ratio for individual galaxies, and the dashed lines represent the sample average for each library. In almost all cases, the effect of the cut is sub-percent. The three panels at the bottom show the age and metallicity of the templates selected by pPXF to fit our spectra, and they correspond to XSL, E-MILES, and sMILES SSP libraries from left to right. The red circles and blue squares are for `full' and `cut' libraries, respectively, where the size of the marker in each panel is proportional to how many times the corresponding template was used. The black dashed lines show the age and metallicity restriction applied to the `cut' libraries. For E-MILES and sMILES, in general, the same templates were selected with similar frequency regardless of which version of the corresponding library was used. As a result, the ratios for individual galaxies (blue and red circles on the top panel) are very close to 1 for almost all galaxies. For XSL, the difference in velocity dispersions from `full' and `cut' libraries are somewhat higher compared to the other two libraries, but still well below our target of 1\%. }
\end{figure*}

We tested the effects of restricting the templates based on stellar age and metallicity (e.g., age between 1 and 12 Gyrs, and metallicity above 0.1 solar), in order to avoid potential non-physical templates. We performed this test on SSP libraries as the age of the stars in stellar libraries is not available from the catalogs in the literature. The results of the tests shown in Figure~\ref{fig:effect_age_metallicity_cut_SSP} demonstrate that in some cases, non-physical templates
(e.g., templates with ages of 15 Gyrs for the XSL SSP library) can be chosen while
using the "full" libraries. 
However, the effect on the inferred stellar velocity dispersion is minimal, well below our target of 1\%, with very few exceptions for the XSL models. We thus conclude that it is not necessary to impose a priori cuts on age and metallicity, even though one should be cautious when interpreting the results. 

\subsection{Templates}
\label{ssec:templates}

\subsubsection{Stellar libraries}

The dominant source of potential systematic errors is the choice of templates. Figures~\ref{fig:comparison_all_stellar_ssp} and~\ref{fig:kcwi_library_comparison} to \ref{fig:nirspec_comparison} show heat maps representing the average difference in stellar velocity dispersion between any pair of combination of stellar templates. For the MUSE, KCWI, and SDSS samples, the difference averaged over the galaxy sample is reported, while for JWST-NIRSpec, the average difference over 21 spaxels is reported. 

It is clear that it is possible to achieve sub-percent agreement among the measurements from the clean stellar libraries, 
irrespective of the galaxy sample or instrument being considered. The full stellar libraries can be significantly off, highlighting the importance of using clean libraries. This is particularly striking in the case of JWST-NIRSpec, where the differences between clean and full stellar libraries can be $\sim$10-15\%.

\subsubsection{Single stellar populations models}
\begin{figure*}
\includegraphics[width=\textwidth]{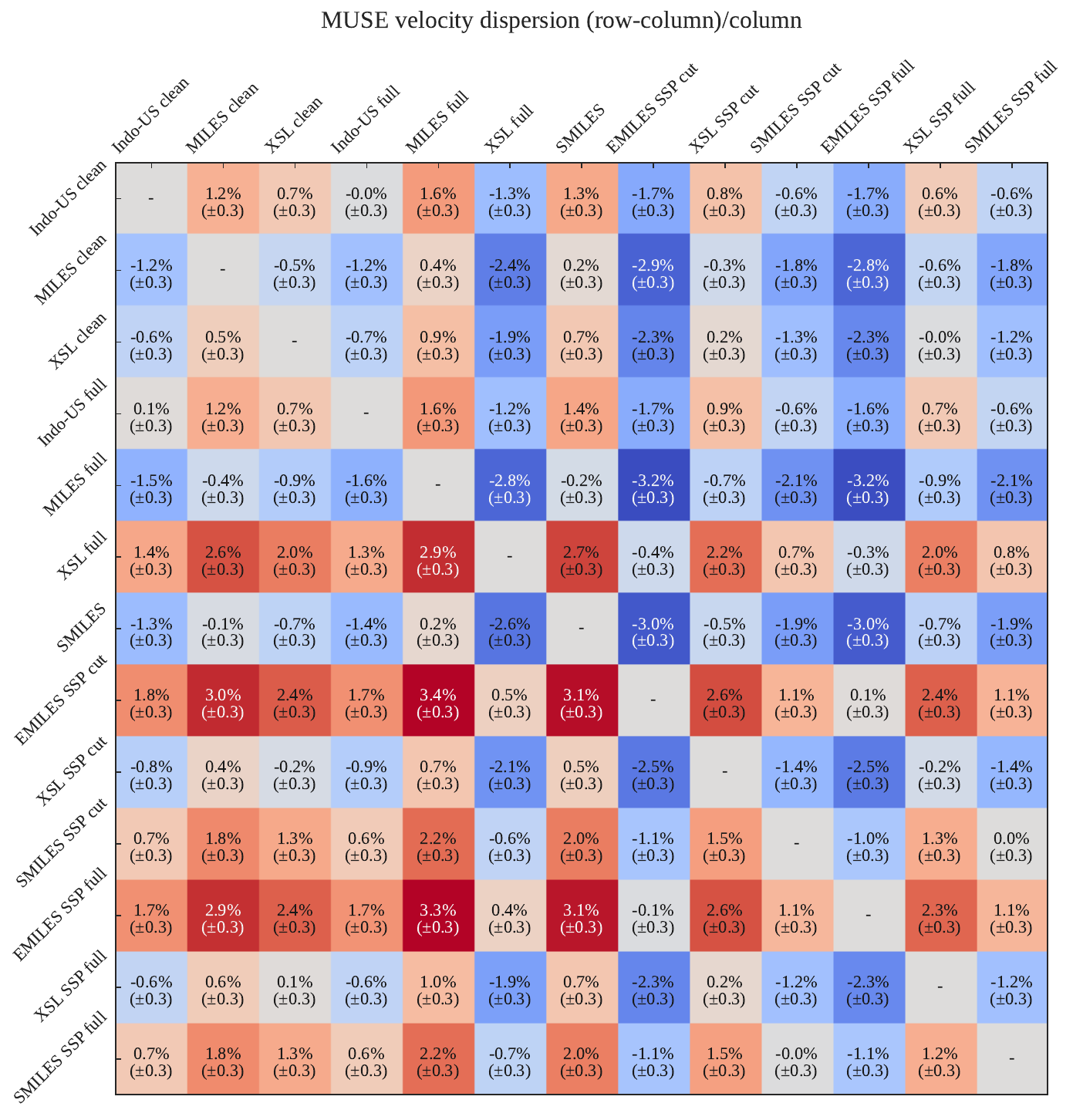}
        \caption{\label{fig:comparison_all_stellar_ssp}
        Comparison of velocity dispersion measurements of the MUSE sample using a range of template libraries. For every pair of libraries, the heat map shows the relative difference in the inferred stellar velocity dispersion, averaged across the sample. The uncertainty is estimated by bootstrap resampling the galaxies. Blue (red) boxes represent pairs in which the library listed on the row yields a lower (higher) value than that listed on the column. The wavelength range is 3600--4500 \AA. Comparisons for the KCWI, SDSS, and JWST/NIRSpec samples are shown in Figures~\ref{fig:kcwi_library_comparison} to \ref{fig:nirspec_comparison}.}
\end{figure*}

As shown in Figures~\ref{fig:comparison_all_stellar_ssp} and~\ref{fig:kcwi_library_comparison} to \ref{fig:nirspec_comparison}, the stellar velocity dispersions based on the SSP models can differ by a few percent from those inferred based on stellar libraries. The difference is more pronounced for the KCWI and SDSS samples than for the MUSE sample. While for the former the SSP libraries give consistently higher stellar velocity dispersions by 2-4\%, the differences are smaller for the MUSE sample. In order to investigate the origin of this discrepancy, we considered a subset of the MUSE sample matched in redshift and stellar velocity dispersion to the KCWI sample. For example, if we restrict the MUSE sample to $z<0.35$ and 220 \kmps $< \sigma < 310$ \kmps, then the stellar vs.~SSP
ratios get higher and closer to what we found for KCWI:
Indo-US vs EMILES SSP: from -1.7$\pm$0.3\% (MUSE-full) to -2.9$\pm$0.8\% (MUSE-restricted), similar to -3.4$\pm$0.4\% obtained for KCWI. The corresponding numbers for MILES vs.~EMILES-SSP are: -2.9$\pm$0.3\%, -3.6$\pm$0.8\%, and
$-2.8\pm0.5$\%. For  XSL vs.~EMILES SSP they are:-2.4$\pm$ 0.3\%, -3.9$\pm$0.8\%, and -4.4$\pm$0.4\%. In all cases, it seems that the SSP ``bias'' is larger when selecting the higher $\sigma$ galaxies at lower redshifts. In general, those will have older ages and higher metallicities than their higher-z and lower-$\sigma$ counterparts, suggesting deficiencies of the SSP libraries in this part of the parameter space. Our sample size is too small to carry out a detailed investigation of the bias between stellar and SSP libraries as a function of galaxy parameters. Hopefully, future studies will address this.

For the time being, we recommend avoiding the SSP models for high precision/accuracy measurements for a number of reasons. First, the SSPs are based on the assumption that spectra are single stellar populations, which is not true in general. Even a linear combination of a few of them might not have sufficient flexibility to describe the star formation history of the galaxy to the level we require. Indeed, as shown above, the SSPs lead to unphysical solutions in some cases, which raises a red flag.  Second, the smaller number of SSP templates compared to stellar templates gives less flexibility to reproduce the diversity in metallicity and elemental abundance. Third, the SSPs are a combination of theoretical and empirical spectra, and we think it is more prudent to stick with empirical spectra for high-precision measurements, given the uncertainties in theoretical atmospheres. 

In conclusion, SSPs are fine if one is interested in obtaining uncertainties at the 2-3\% level. However, we find that they are not appropriate for the sub-percent accuracy required for cosmology.

\subsection{Wavelength range}
\label{ssec:wave}

The wavelength range used for the fit is often dictated by practical considerations such as instrumental setup, wavelength coverage of the templates, and the need to avoid features from contaminants (\ref{ssec:JWST}). Within these constraints, it is best to choose a wavelength region that contains deep absorption features such as CaHK, the G-band, and the \ion{Ca}{II} Triplet. As we discussed above, the choice of polynomial orders to describe multiplicative and additive continuum terms depends on the wavelength region chosen (see Section \ref{ssec:polynomials}). Fortunately, our extensive tests show that once appropriate polynomial orders are selected and clean stellar libraries are employed, the specific choice of the spectral region is a subdominant contribution to the systematic error budget. Small changes in wavelength coverage (i.e., shifts by 100 \AA\ or so) are absolutely negligible for the quality of the data presented here. Even larger changes make a subdominant contribution to the error budget \citep[see also][for a similar result]{DAGO23}. 

\begin{figure}
    \includegraphics[width=\columnwidth]{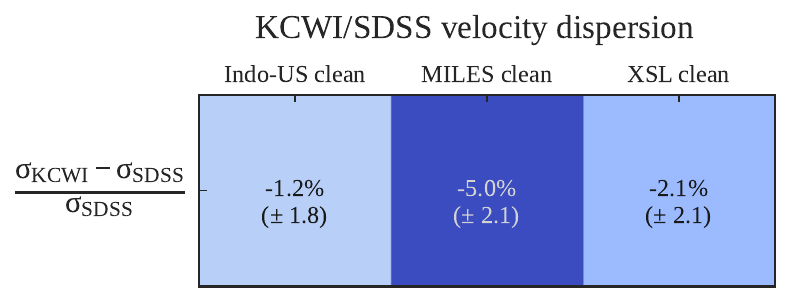}
    \caption{Comparison of the velocity dispersion measured from the KCWI and SDSS spectra of the same galaxies, using the three ``clean'' stellar libraries. Note that SDSS and KCWI cover two different wavelength ranges (3600--5350 \AA\ and 3600--4500 \AA, respectively). The KCWI data have been convolved to match the SDSS seeing and integrated within the SDSS fiber for a fair comparison. Though the difference in the case of MILES is relatively large, we stress that SDSS data for these specific galaxies are not of sufficient quality to measure stellar velocity dispersions with precision and accuracy better than 5\%. \label{fig:KCWI_SDSS_comparison}}
\end{figure}

We took advantage of the SDSS spectra available from the SDSS database for the 14 galaxies with KCWI spectra to verify the effects of changing significantly the wavelength range. The wavelength ranges we adopted vary substantially (3600-4500 \AA\ for KCWI and 3600-5350 \AA\ for SDSS), with the SDSS including important redder features such as the MgB triplet and Fe5270. We could not go redder with SDSS because the XSL library is not sufficiently well behaved across the dichroic for our target level of precision and accuracy. To ensure a fair comparison, following \citet{Knabel24}, we convolved the KCWI data with a kernel to match the SDSS seeing and then measured the stellar velocity dispersion within a circular aperture of diameter $3\arcsec$. The SDSS error dominates the uncertainty, since the SDSS S/N is substantially lower than KCWI. The results are shown in Figure~\ref{fig:KCWI_SDSS_comparison}. For the Indo-US and XSL clean libraries, the stellar velocity dispersions agree within one $\sigma$, while for MILES, the discrepancy is 2.5$\sigma$. Given the uncertainties, we see no significant evidence to support a difference, although the uncertainty is relatively large. 

This comparison should be taken with caution and should be repeated with better quality data, ideally with a larger sample.  \citet{Knabel24} discussed this comparison in more detail. The main conclusion is that the SDSS data for these specific galaxies are not of sufficient quality to measure stellar velocity dispersions with precision and accuracy better than 5\%, confirming the initial assessment by \citep{Bolton08}.  

\citet{Shajib25} presented a comparison between kinematics obtained with JWST in the \ion{Ca}{II} Triplet region and with KCWI in the blue part of the spectrum. \citet{Knabel24}  presented  further comparisons between SDSS and KCWI. 

In our analysis, we chose not to discuss the wavelength region much further. However, in our suggested recipe, we include a reminder to check for the effects of the wavelength range.

In summary, we consider the following rest-frame spectral ranges. For MUSE and KCWI, we adopted the range 3600-4500 \AA\ , where the quality of the template libraries sets the blue cutoff and the red cutoff is imposed by the redshift of the targets and the instrumental setup. For JWST, we used the range 8303-8838 \AA\, for reasons discussed in Section~\ref{ssec:JWST}. For SDSS, we considered 3600-5350 \AA, taking advantage of the redder coverage to increase the information content of the spectra and avoiding complications from the UVB-VIS dichroic in the XSL templates.

\section{A recipe for accurate stellar velocity dispersions}
\label{sec:recipe}

This section describes a recipe to quantify and minimize residual systematic errors and covariance. The recipe consists of a number of steps that will be summarized at the end of the section. At the core of the recipe is a well-established statistical method to marginalize over systematic uncertainties based on information contained in the data themselves. This section is focused on statistical details, followed by a presentation of the final end product described in Section~\ref{sec:results} and in Table~\ref{table:results}, where we summarize the results obtained with our method. We also include results we would obtain by the simplest approach, namely, a straight average of the fits obtained with the three clean stellar libraries.

In practice, we consider the possible choice of template library as a nuisance parameter and marginalize over it weighted by the probability of the model given the data.  However, since the data and model parameters vary, to obtain a fair marginalization we cannot use a simple likelihood ratio, but we need to use a more general metric. Following, for instance, \cite{Birrer19b}, we assigned the statistical weight, $w$, based on the Bayesian information criterion (BIC), defined as \citep{Schwarz1978}:

\begin{equation}
    {\rm{BIC}} = k\,{\rm{ln}}(n) - 2\, {\rm{ln}}(\hat{L}) ,
\label{eq:BIC}
\end{equation}

\noindent where $k$ is the number of free parameters in the model, $n$ represents the number of data points (or pixels) used in the fit after masking, and $\hat{L}$ stands for the maximum value of the likelihood function, corresponding to the model. 

When the template libraries have been properly cleaned, the resulting likelihoods from fits to a single spectrum will generally not distinguish a preferred library. Instead, we determine the combined BIC for a sample of galaxies or for a set of spatially binned spectra for a single galaxy observed with IFU spectroscopy. Note that for a sample of galaxies or set of bins, we have to construct the BIC for the whole sample since the first term of the right-hand side of \autoref{eq:BIC} is non-linear in $n$. In other words, if there are $N$ galaxies in the sample with $n_i$ data points each, and they are each fit with $k_i$ degrees of freedom  resulting in likelihood $\hat{L_i}$  the equation becomes
\begin{equation}
    {\rm{BIC}} = \left(\sum_{i} k_i \right)\,{\rm{ln}}\left(\sum_i n_i \right) - 2\, \sum_i {\rm{ln}}\,\hat{L_i} .
\label{eq:BIC-sample}
\end{equation}
The likelihood term is effectively a $\chi^2$ with an important caveat. Standard practice in kinematics fits is to scale the noise levels by a multiplicative factor so that the $\chi^2$ per degree of freedom is one, in order to account for potential error under/overestimates. We adopt this practice, but we are careful to adopt the same multiplicative factor for a given galaxy across all our tests, to retain the information provided by the difference in $\chi^2$ across fits.

All of the methods we present are equally valid for the case of a sample of galaxies and for the case of a set of spatial bins measured across a single galaxy. In the following text, where we refer to a sample of galaxies, we could also refer to the spatial bins of a single galaxy. In general, for spatially resolved samples like KCWI and MUSE, we would prefer to treat each object individually. However, none of the objects in our KCWI and MUSE samples have sufficient S/N across a large enough number of spatial bins to be able to strongly distinguish a BIC-preferred template library given the uncertainties in the BIC comparisons. For this reason, and the reasons outlined in Section~\ref{sec:galaxies}, we integrate the datacubes to achieve the highest possible S/N for each object in the sample. Therefore, in the results described in Section~\ref{sec:results}, the preference for a specific template library is driven by the properties of the sample as a whole, including properties specific to the instrument used to obtain the data. However, we do not consider any of these BIC preferences to be universally applicable to data obtained with that instrument, nor to the same sample of galaxies observed with another instrument. Complete studies of spatially resolved datasets should also be assessed at the level of spatial bins. The procedures will be exactly the same as outlined in this section.

\subsection{Counting free parameters}

The definition of the number of free parameters in the BIC is non-trivial and requires some discussion.
The number of free model parameters, $k$, is given by $k=M+N_{\rm poly}+N_{\rm temp}$, where the M is typically 2 and accounts for the stellar velocity dispersion and line-of-sight velocity \citep[but it could be more if there are multiple components, see, e.g.,][]{Shajib25}, $N_{\rm poly}$ is the number of corrective polynomial orders (both additive and multiplicative), and $N_{\rm temp}$ is the number of ``significant templates,'' namely, the number of nonzero templates, or in practice the number of templates that contribute above a certain threshold (in our case 1\%) to the fit. The threshold is needed for two reasons. First, stellar libraries can include a large number of spectra that are not relevant to fit the data (e.g., B stars) and those should not be counted in order to obtain a fair representation of the effective freedom of the fit.

Second, from a numerical standpoint, the rationale for omitting the weights of zero (or nearly zero) templates when estimating free parameters is that when a template weight reaches the boundary at the optimal fit, the positivity constraint in the template fit behaves like an equality constraint. This effectively eliminates one free parameter.
During the constrained quadratic template optimization, pPXF identifies the optimal weights by aiming to minimize the $\chi^2$ to machine precision. This process can result in some very small weights, which, while formally necessary to optimize $\chi^2$, make no significant contribution to the fit from a physical perspective.

The count of effective degrees of freedom attributable to the templates used in the fit is therefore paramount to the proper comparison of model BICs. Intuitively, the threshold is related to the noise level of the galaxy spectra. For our data quality S/N of~30--160 \AA$^{-1}$ (i.e., 1-3\% per \AA), we find that 1\% is the appropriate threshold using the following procedure. We first carry out a pPXF fit with all the templates in the library, and then we repeat the fit by keeping the spectra that contribute only above a certain threshold to the initial fit. As expected, we find that for a high threshold (10-20\%), the BIC is large because there are not enough templates to obtain a good fit. Conversely, below $\sim1$\%, the $\chi^2$ does not improve enough to compensate for the increased number of degrees of freedom in the BIC. This threshold depends on the quality of the data, approximately in a manner inversely proportional to the S/N. We recommend following our procedure to set the threshold for each specific dataset. \citet{Thomas_Lipka22} offered an alternative approach to counting the effective number of free parameters.

\subsection{Computing the weights}

In the following formulas, we use subscripts $k$ to label the models (e.g., template libraries) that are being compared, while $i$ refers to the sample of galaxy spectra or spatial bins. The probability of a model, $F_k$, given the data is formulated as

\begin{equation}
\begin{split}
    P(F_k | \bm{d} ) &\propto P(\bm{d} | F_k) P(F_k) \\ 
    &\approx \exp \left[ -\frac{1}{2} \mathrm{BIC}_k \right] P(F_k).
\label{prob_of_model}
\end{split}
\end{equation}

\noindent Hence, the relative probability between two fits, $F_1$ and $F_2$, given the data, can be written in terms of the fits' corresponding BIC values,

\begin{equation}
    \frac{P(F_1 | \bm{d})}{P(F_2 | \bm{d})}  = \rm{exp} \left[ -\frac{1}{2} (\rm{BIC_1 - BIC_2}) \right].
\label{prop_of_BIC_models}
\end{equation}

However, we only have a noisy estimate of the BIC, with the noise given by statistical noise and sampling noise. Thus, we need to define the evidence ratio function $f(x)$ in terms of a dummy variable $x$ sampling the ``true'' difference between a fit's corresponding BIC value and the lowest BIC among all possible fit configurations as discussed above, BIC$_{\rm{min,k}}$,

\begin{equation}
    f(x) \equiv
    \begin{cases}
    1 &, x \leq 0,\\
    {\rm{exp}} \left[ -\frac{1}{2} (x) \right] &, x > 0.
    \end{cases}
\label{evidence_ratio_function}
\end{equation}

To find a model's associated statistical weight, $w_k$, we marginalized over $x$, with the probability of $x$ given the measured $\Delta {\rm BIC}_k$ described by a Gaussian, $g$, centered on $\Delta$BIC$_k$ with a variance of $\sigma_{\Delta {\rm BIC},k}^2$. size. In the absence of uncertainty, $x$ would be strictly positive. However, when convolving for the effect of the uncertainty, the width of the convolution kernel (set by $\sigma_{\Delta {\rm BIC},k}^2$) might extend some part of the kernel’s tail to $\Delta {\rm BIC}_k < 0$, which we allow.

The standard deviation, $\sigma_{\Delta {\rm BIC},k}$, represents the uncertainty in the $\rm{\Delta} BIC$ calculation and is dominated by numerical effects and population sampling. There could be, for example, one galaxy with unusual abundance ratios that is the best fit by a lucky template and skews the BIC. In contrast to the strong lensing cases described by \cite{Birrer19b} and \cite{Shajib22b}, convergence is not an issue for the kinematic fits described here. The BIC uncertainty can then simply be estimated by bootstrap resampling the sample of galaxies. Importantly, there is covariance between the BIC calculated using a pair of stellar libraries: some galaxies might be better fit by more than one template library than others, and therefore, we need to compute the scatter in $\Delta$BIC by bootstrap resampling the sample of galaxies, rather than computing the scatter independently for each library. Typically, we find $\sigma_{\Delta \rm{BIC}}=100-200$ for the samples under consideration here, although it is clear that the BIC and its uncertainty depend on sample size.

From the bootstrapped $\rm{\Delta} BIC$ distributions for the SDSS sample, we test the null hypothesis that they are derived from the Gaussian distribution $g$. Using $\rm{\Delta} BIC_k$ for the three clean libraries (Indo-US, MILES, and XSL), we found Kolmogorov-Smirnoff statistics of 0.021, 0.018, and 0.017, and p-values of 0.75, 0.91, and 0.93, indicating that the null hypothesis is not rejected and the distributions are reasonably approximated by the Gaussian, $g$.

The marginalization over $x$ is given by the integral:

\begin{equation}
    w_k = \int f(x) \frac{1}{\sqrt{2 \pi} \sigma_{\Delta {\rm BIC},k}} \exp \left[-\frac{1}{2}\left(\frac{x - \Delta {\rm BIC}_k}{\sigma_{\Delta {\rm BIC},k}} \right)^2\right] {\rm d}x.
\label{BIC_weight}
\end{equation}

The integral can be simplified analytically to gather some insight and for numerical precision. We set for simplicity of notation $\Delta \coloneqq \Delta {\rm BIC}_k$ and $\sigma \coloneqq \sigma_{\Delta \rm{BIC},k}$. The part for $x\leq0$ is simply the integral of a Gaussian, simplified by setting $y=(x-\Delta) / \sigma$, while the part for $x>0$ can be solved 
by setting $z=(x - \Delta)/\sigma + \sigma/2$ to obtain 

\begin{equation} \label{BIC_weight2}
    w_k = \int_{-\infty}^{-\Delta /\sigma} \frac{e^{-y^2/2}}{\sqrt{2 \pi}} \mathrm{d}y  + \exp\left[\frac{\sigma^2}{8} -\frac{\Delta}{2}\right] \int_{-\Delta/\sigma+\sigma/2}^{\infty} \frac{e^{-z^2/2}}{\sqrt{2 \pi}}  \mathrm{d}z.
\end{equation}

\noindent
If $\Delta$ is much larger than $\sigma$ (i.e., if a BIC is significantly worse than the best BIC), then the first term becomes negligible, and the second term behaves as an exponential down-weight. 

Having defined the weights, the mean $\bar{\sigma}$, systematic uncertainty $\Delta \bar{\sigma}$, and statistical uncertainty $\delta \bar{\sigma}$ of the velocity dispersion of an individual galaxy is given by the standard expressions

\begin{equation}
\bar{\sigma} = \frac{\sum_{k} w_k \sigma_k}{\sum_k, w_k}
\label{eq:mean}
,\end{equation}
\begin{equation}
(\Delta \bar{\sigma})^2 = \frac{\sum_{k} w_k (\sigma_k-\bar{\sigma})^2}{\sum_k w_k},
\label{eq:sys_error}
\end{equation}
and
\begin{equation}
\delta \bar{\sigma} = \sqrt{\frac{\sum_{k} w_k (\delta \sigma_k)^2}{\sum_k w_k}},
\label{eq:stat_error}
\end{equation}
where the index $k$ runs through the choices (e.g., template library), and $\delta \sigma_k$ is the statistical uncertainty estimated by pPXF. For an individual system, the total error is given by the quadratic sum of the systematic and statistical errors.

When the number of template libraries is small (as in the case considered here), it is prudent to adopt Bessel's correction to get the unbiased estimators (see Appendix~\ref{app:Bessel} for a derivation)

\begin{equation}
(\Delta_B \bar{\sigma})^2 = \frac{ \sum_{k} w_k (\sigma_k-\bar{\sigma})^2}{\sum_k w_k - \sum_k w_k^2 / \sum_k w_k}.
\label{eq:sys_error_Bessel}
\end{equation}

For a sample of galaxies or multiple spatial bins within a galaxy, in addition to measuring the average the mean $\bar{\sigma}_i$, we can estimate 
the covariance matrix as

\begin{equation}
C_{ij} \bar{\sigma} = \frac{\sum_{k} w_k (\sigma_{k,i}-\bar{\sigma}_{i})(\sigma_{k,j}-\bar{\sigma}_{j}) }{\sum_k w_k} + \delta_{i,j} \delta \bar{\sigma}_{i}^2,
\label{eq:covariance}
\end{equation}
where the indexes $i,j$ identify the elements of the covariance matrix, and $\delta_{i,j}$ is the Kronecker's delta function. Along the diagonal, the first term is the systematic error, and the second is the statistical error.

For a small number of stellar libraries, again, we recommend Bessel's corrected estimator, which is expressed as
\begin{equation}
C_{B,ij} \bar{\sigma} = \frac{ \sum_{k} w_{k} (\sigma_{k,i}-\bar{\sigma}_{i})(\sigma_{k,j}-\bar{\sigma}_{j}) }{\sum_k w_k - \sum_k w_k^2 / \sum_k w_k} + \delta_{i,j} \delta \bar{\sigma}_{i}^2.
\label{eq:covariance_Bessel}
\end{equation}

\subsection{Verifying polynomial orders setting}

Our tests show that the BIC does not vary sufficiently to distinguish polynomial orders. This is understandable in terms of mathematical degeneracies: within plausible ranges, all polynomial orders give good fits. However, if the polynomial orders are too low or too high, they will artificially enhance the discrepancy between the results of applying different templates. Therefore, it is important to choose a combination of polynomial orders in the range where the stellar velocity dispersion is stable and the scatter between templates ($\Delta \bar{\sigma}/\bar{\sigma}$) is smallest. 

For a sample of galaxies or multiple spatial bins within a galaxy, identified by the subscript $i$, we would have to compute the average $\sigma_k$ across the sample $<\sigma_k>_i$, then compute the weighted average of it across templates index $k$ using the global BIC,

\begin{equation}
\bar{<\sigma>_i} = \frac{\sum_{k} w_k <\sigma_k>_i}{\sum_k, w_k},
\label{eq:meanmean}
\end{equation}
and the systematic uncertainty on the mean,

\begin{equation}
(\Delta {<\sigma>_i})^2 = \frac{\sum_{k} w_k (<\sigma_k>_i-\bar{<\sigma>_i})^2}{\sum_k w_k}.
\label{eq:sys_error_mean}
\end{equation}

For a small number of stellar libraries, as in the case discussed here, we recommend applying Bessel's correction,

\begin{equation}
(\Delta_B {<\sigma>_i})^2 = \frac{ \sum_{k} w_k (<\sigma_k>_i-\bar{<\sigma>_i})^2}{\sum_k w_k - \sum_k w_k^2 / \sum_k w_k}.
\label{eq:sys_error_mean_Bessel}
\end{equation}

\begin{figure*}
\includegraphics[width=\textwidth]{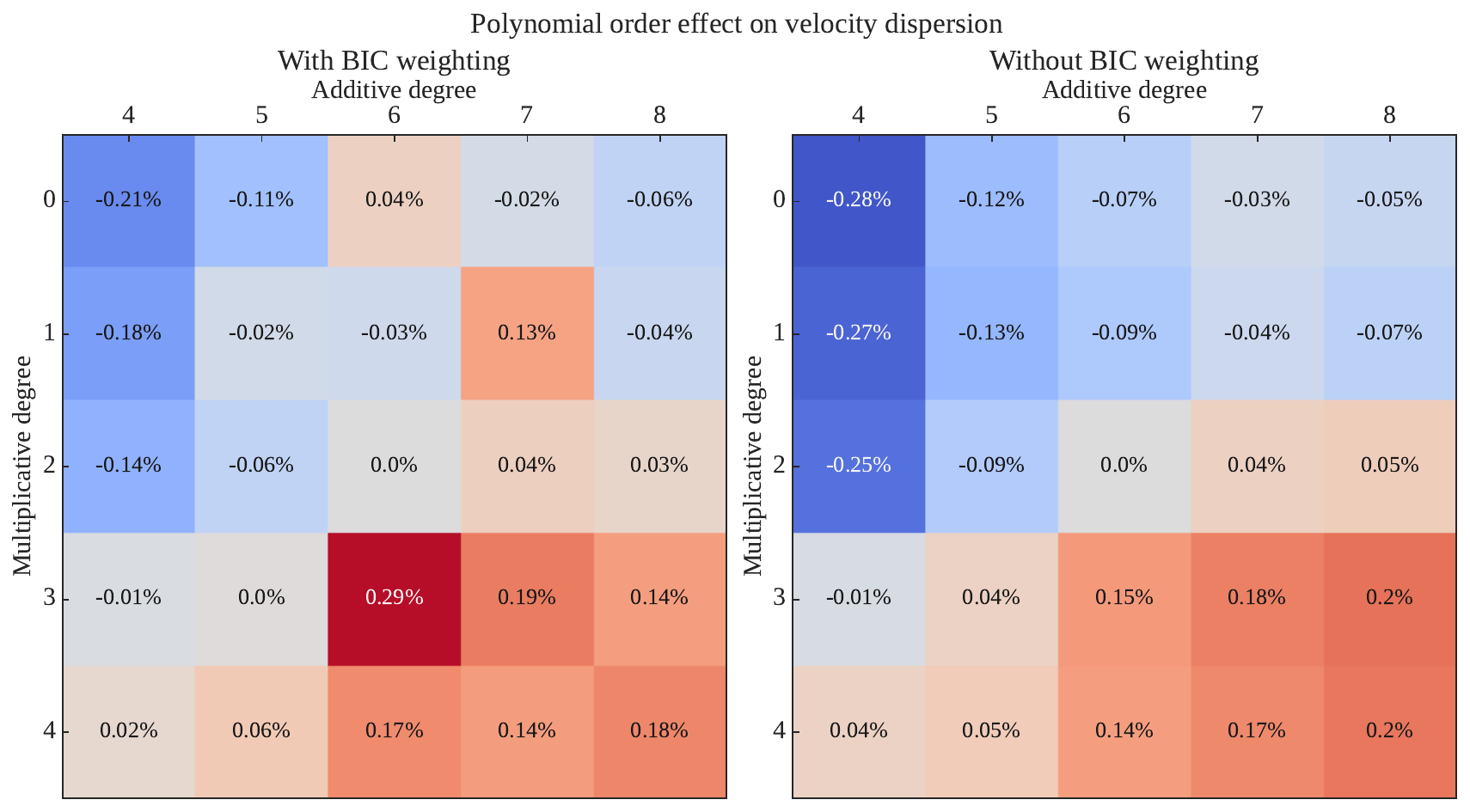}
        \caption{\label{fig:polynomial_order}
    Dependency of the average stellar velocity dispersion inferred for the MUSE sample as a function of polynomial orders, relative to the fiducial setting. The variations are much smaller than 1\%, with and without BIC weighting. }
\end{figure*}

\begin{figure*}
\includegraphics[width=\textwidth]{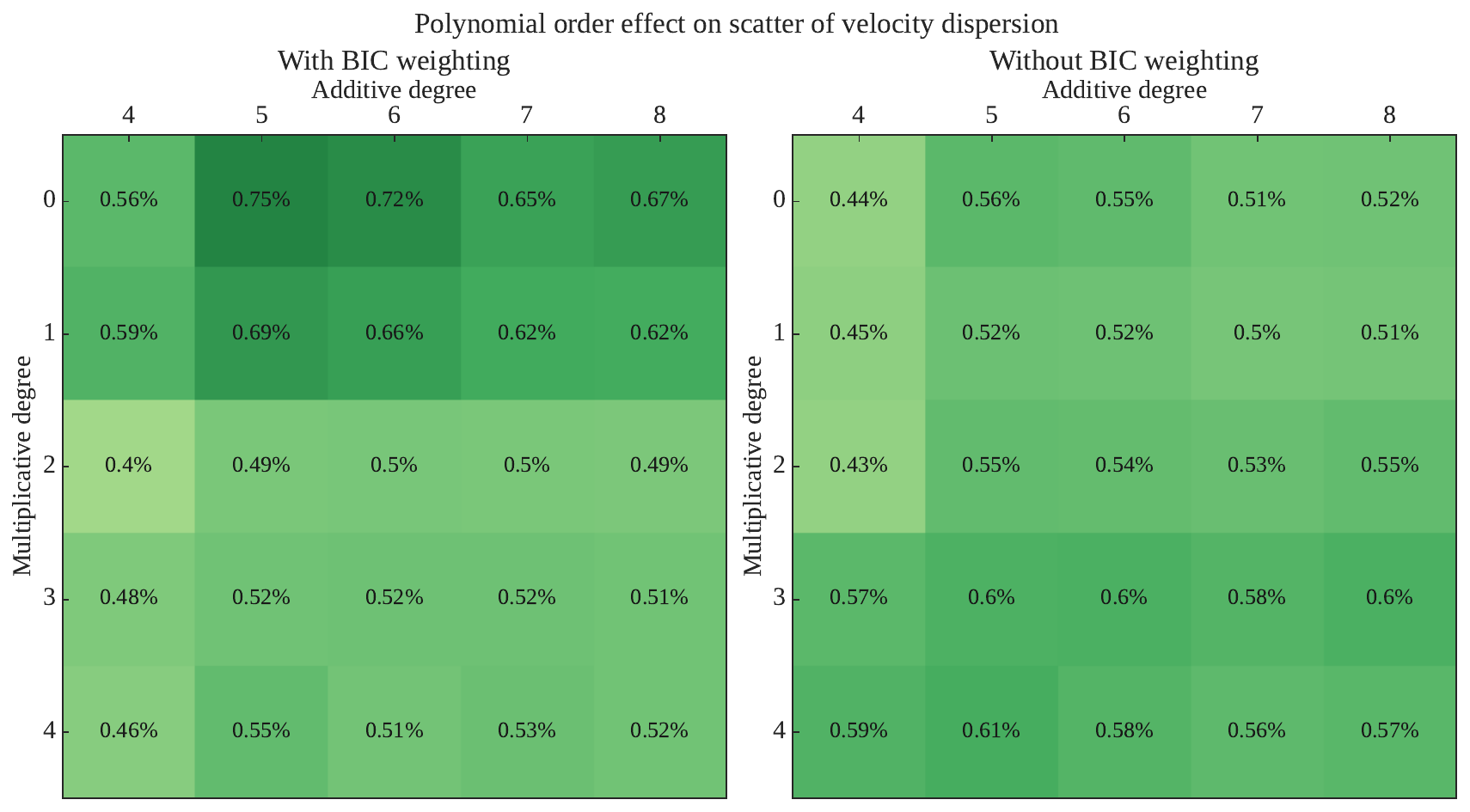}
        \caption{\label{fig:polynomial_order2}
    Average Bessel-corrected systematic scatter of the stellar velocity dispersions for the MUSE sample as a function of polynomial orders, normalized to the weighted average velocity dispersion of the sample. The variations are much smaller than 1\%, with and without BIC weighting.}
\end{figure*}

In the end, it is important to verify that the combination of polynomial orders  
$\Delta <\sigma>_i/\bar{<\sigma>_i}$ is around the minimum within the uncertainty.
In Figures~\ref{fig:polynomial_order} and~\ref{fig:polynomial_order2}, we show as an example the dependency of the inferred stellar velocity dispersion as a function of polynomial orders for the MUSE sample. The left-hand side panel shows the BIC-weighted average of the three clean stellar libraries, while the right-hand side panel shows the straight average. In all cases, the results change much less than 1\% for polynomial orders around our fiducial values. Bessel-corrected estimators have been used.

\subsection{Summary of the recipe}

We summarize our suggested recipe step by step to obtain accurate and precise stellar velocity dispersion below. 

\begin{enumerate}
\item Use data with sufficient S/N. If the S/N is too low, the inference will not only be more noisy, but can also be biased. The precise threshold depends on the wavelength coverage and spectral resolution of the dataset and also on the target accuracy and precision. This can be estimated by simulations or checked post facto using this recipe itself, varying the threshold in S/N.
\item Use clean stellar template libraries. Popular stellar libraries include spectra that might be interesting for other applications, but have insufficient quality for precise and accurate determination. We provide our ``clean'' list for MILES, XSL, and Indo-Us through GitHub, with the caveat that additional cuts and masking might be required depending on the wavelength and quality of the data under consideration. 
\item Run initial tests on additional and multiplicative polynomial orders to identify the range in which the results are stable. Depending on the wavelength range and quality of the data, it might be necessary to marginalize over polynomial orders while computing the total systematic error budget.
\item Measure the stellar velocity dispersion with all the high-quality (clean) stellar libraries that have sufficient resolution and wavelength coverage for the data set. 
\item Compute the Bayesian information criterion to evaluate the relative goodness of fit of the data for each of the template libraries.
\item Estimate the systematic error associated with template libraries, and correlation between multiple galaxies or spatial bins, using the equations provided in this section. 
\item In this optional step, if the systematic error associated with the template library is smaller than 0.5\%, and a better estimate of the systematic error is needed,  the additional term is computed by evaluating residual scatter by marginalizing over polynomial orders and small changes in the wavelength range.
\end{enumerate}

\section{Results}
\label{sec:results}

In this section, we summarize and discuss our main results, namely, the precision and accuracy of the stellar velocity dispersion measurements achieved by applying our recipe to our datasets. We stress that, while the recipe is general and can be applied to any dataset, the precision and accuracy that can be achieved are specific to each dataset and depend on the S/N, resolution, wavelength range, and intrinsic properties of the galaxy spectra under consideration. Therefore, we first discuss each dataset separately and  draw some general conclusions.
The key numbers are summarized in Table~\ref{table:results}. For convenience, we report both the values obtained with our optimal recipe and also those obtained with a straight average of the three clean stellar libraries. 

\subsection{MUSE}  

For the MUSE sample, we find the average statistical uncertainty to be of order 2\%. This is not surprising, given the quality of the dataset. We find that the data prefer the Indo-US library over the XSL library and strongly disfavor the MILES library. We note that this preference is made possible by the large dataset. If we had a single galaxy, the difference in BIC would be much less statistically significant.

The BIC-weighted average systematic error  is much smaller than our target of 1\%, even with Bessel correction. This is, of course, only the error associated with template choice. Other potential sources of error at the level of 0.2-0.5\% are the wavelength range, instrumental resolution, and polynomial order, and those should be investigated if we want to push the accuracy to the level of 0.3-0.4\%. 

Not surprisingly, the  velocity dispersion has positive covariance across the sample: templates that yield higher (lower) velocity dispersion do so consistently for all galaxies. Importantly, the amplitude of the covariance is also well below 1\%. If we were to ignore the information from the BIC and weigh the three clean stellar libraries equally, the uncertainties associated with the stellar template would increase to the values indicated in the lower part of Table~\ref{table:results}. The increase is noticeable and strengthens the importance of adopting our proposed recipe. However, even with this suboptimal weight, the systematic errors and off-diagonal elements of the covariance matrix remain on the sub-percent level.

\begin{table*}
\caption{Summary of results.}
\begin{center}
\begin{threeparttable}
\resizebox{\textwidth}{!}{
\begin{tabular}{l c c c c c c c c} 
    \hline
Sample    & $<\delta \bar{\sigma}/\bar{\sigma}>$ & $<\Delta \bar{\sigma}/\bar{\sigma}>$ & $\sqrt{<C_{i,j}\bar{\sigma}/\bar{\sigma_i}\bar{\sigma_j}>_{i\neq j}}$ & $<\Delta_B \bar{\sigma}/\bar{\sigma}>$ & $\sqrt{<C_{B,i,j}\bar{\sigma}/\bar{\sigma_i}\bar{\sigma_j}>_{i\neq j}}$ & Indo-US & XSL & MILES \\
\hline
MUSE-BIC     &   2.04\%  &  0.21\%   &  0.12\%  &  0.78\%  &  0.45\%   & 0.96  & 0.04  & 0 \\ 
KCWI-BIC     &   0.86\%  &  0.02\%  &  0.01\%   &  0.67\%   &  0.47\%  & 0.9995 & 0 & 0.0005 \\
NIRSpec-BIC  &    5.73\%  &   0.60\% &  0.13\%   & 0.85\%  & 0.19\%    & 0.47   & 0.53   & --   \\
SDSS-BIC     &   3.24\%  &   0.01\%  &  0.01\%   &  1.46\%  & 1.27\%   &  0.00001     & 0.99998  & 0.00001\\
\hline
MUSE         &   2.04\%  &  0.78\%   &  0.47\%   &  0.95\%  &  0.58\%  & 0.333 & 0.333 & 0.333 \\ 
KCWI         &   0.90\%  &  0.79\%  &   0.70\%  &  0.97\%  &  0.86\%  & 0.333 & 0.333 & 0.333 \\
NIRSpec      &   5.74\%  &   0.60\% &  0.14\%   & 0.85\%  & 0.19\%    & 0.5   & 0.5   & --   \\
SDSS         &   3.30\%  &   1.30\%  & 1.15\%  & 1.60\% &  1.40\%  & 0.333 & 0.333 & 0.333 \\
\hline
\end{tabular}
}
\end{threeparttable}
\end{center}
\label{table:results}
\tablefoot{
For each dataset, we report the average statistical error, the average systematic uncertainty associated with the choice of clean template library, the average amplitude of the off-diagonal terms of the covariance matrix between elements of the sample, the Bessel-corrected average systematic uncertainty and off-diagonal covariance matrix, and the weights given to the three libraries. The top part of the table reports the results obtained by weighting the three clean stellar libraries according to the BIC as described in Section~\ref{sec:recipe}. The bottom part of the table reports the results based on equal weights.  See text for details.
}
\end{table*}

\subsection{KCWI}

The results for KCWI are very similar to those obtained for MUSE, with the main difference being that Indo-US is now more strongly favored with respect to both XSL and MILES.
This might be due to the superior S/N of the data, reflected in the sub-percent statistical errors. Since the data inform us that they clearly prefer Indo-US, then the residual systematics obtained with our procedure are vanishingly small and a Bessel correction is crucial to obtain a realistic estimate of the systematic error. 
For comparison, a straight average of the three clean stellar libraries leaves systematic residual uncertainties and covariance below 1\%. 

\subsection{NIRSpec}

For the NIRSpec data, we find sub-percent systematic errors and off-diagonal elements, similar to all the other datasets considered in this paper. The average random uncertainty is $\sim$6\% per spaxel. Although this level is higher than that of other samples, this is driven by the somewhat lower signal-to-noise ratio of the single spaxels in the NIRSpec data, and by the QSO contamination. The random uncertainty can be decreased by summing over multiple spaxels, albeit at the loss of spatial resolution. BIC-weighting prefers XSL over Indo-US only marginally.

\subsection{SDSS}

The results for SDSS are qualitatively in line with the other instruments, even though the accuracy and precision is lower, as expected given that the S/N of the data is significantly lower. As expected, average random errors are higher than MUSE and KCWI, between 3-4\%. Remarkably, even with the lower data quality, systematic errors and covariance are below 1.5\% for BIC weighting and still below 2\% with the straight average. The most significant difference is that the SDSS spectra prefer XSL over Indo-US, whereas the opposite is true for MUSE and KCWI. MILES is always disfavored. Our suggested explanation for this difference is that the longer and redder wavelength coverage of the SDSS spectra compared with MUSE and KCWI (3600-5350~\AA\, vs 3600-4500~\AA). For example, the Indo-US flux calibration might be better in the blue but less accurate than XSL in the longer and redder wavelength range. The stellar populations that dominate the redder absorption features could also be better described by the XSL library. In contrast, those that dominate in the blue could be better described by Indo-US. This difference highlights the importance of running all the clean stellar libraries. 

\subsection{General results}

We now summarize results that are general and valid across the samples considered here. We also expect them to be valid for other samples with data of comparable quality. First, once properly cleaned stellar libraries are considered, residual systematics on stellar velocity dispersion due to the template choices are sub-percent. Second, the uncertainty can be further reduced by taking advantage of the information content of the data to select which set of templates better describes the data, according to the procedure described in Section~\ref{sec:recipe}. We note that the exact outcome of this process will depend on the data quality and quantity (that set the statistical significance) but also on the wavelength range covered by the data and by the stellar populations in the target galaxies (that might be best described by a specific library or not).     
Third, the covariance between different galaxies or different spatial bins is also at the sub-percent level.
Fourth, there is no single clean stellar library that is preferred by all datasets, so it is important to use all three.

We carried out additional tests of our proposed recipe with mock samples constructed from the clean stellar template libraries of Indo-US and XSL, demonstrating that the method is able to accurately recover the true velocity dispersion to sub-percent accuracy. The details of the mock sample and results are included in Appendix~\ref{app:mock_data}.

\section{Summary}
\label{sec:summary}

We carried out a detailed study of systematic uncertainties associated with stellar velocity dispersion measurements in samples of massive elliptical galaxies at redshift $z\sim0.1-0.65$. We used spectra with S/N$\sim$30-160 \AA$^{-1}$ obtained with three different state-of-the-art instruments: VLT-MUSE, Keck-KCWI, and JWST-NIRSpec. For comparison with previous works, we also studied a sample of relatively high-S/N spectra from SDSS (S/N$>$15 \AA$^{-1}$). We consider three state-of-the-art empirical stellar libraries (Indo-US, MILES, and XSL) as templates. For completeness, we also investigated synthetic libraries that are $\alpha$ enhanced (sMILES) and composite stellar population synthesis (SSP) models that have been used in previous studies. Our main results can be summarized as follows:

\begin{enumerate}
\item For the quality considered here, the main source of systematic uncertainty is the choice of the stellar library to be used as a kinematic template.
\item It is very important to use ``clean'' stellar libraries, that is, to remove spectra that have defects, such as gaps, poor correction of slit losses, atmospheric absorption,  differential diffraction, and so on. Improperly cleaned stellar libraries can result in differences of order a few percent and sometimes up to 10-15\%. 
\item We find that SSP and synthetic $\alpha$-enhanced libraries can produce results that differ by a few percent with respect to clean stellar libraries and we discourage their use for high-precision stellar kinematics. For the SSP libraries, we interpret the bias as being due to incomplete coverage of parameter space. This yields insufficient flexibility to reproduce the detailed star formation and chemical enrichment history of massive elliptical galaxies to our desired level of precision. For the $\alpha$-enhanced libraries, the bias is likely due to known limitations in the modeling of stellar atmospheres \citep{Knowles23}.
\item Once clean stellar libraries are considered, the residual uncertainty related to the choice of templates is sub-percent for all the primary datasets considered here (MUSE/KCWI/NIRSpec). Even for the lower S/N SDSS spectra they are below 2\%. The off-diagonal elements of the covariance matrix between galaxies in a sample and between spatial bins within a galaxy are positive and also sub-percent.
\item We suggest a procedure that uses the Bayes information criterion (BIC) to quantify how well each stellar library matches the dataset and assign weights based on the difference in BIC between the stellar libraries. With this optimal weighting, the residual systematic uncertainty and covariance are further reduced to levels well below 1\%.
\item The weights of the clean stellar libraries depend on the sample. Our KCWI and MUSE data prefer the clean Indo-US library, while the JWST and SDSS spectra prefer the clean XSL library. It is thus important to make the measurements with all three clean libraries to obtain a complete picture.
\item Other sources of residual uncertainty related to the orders of the polynomials used to model the continuum are shown to be on the order of 0.2-0.5\% for our dataset. 
\end{enumerate}

In conclusion, we have shown that adopting the method suggested here leads to a substantial improvement in the precision and accuracy of stellar kinematics measurements, as well as the estimation of the residual uncertainties. While the final precision and accuracy will depend on the properties of the dataset (including the wavelength range), the methodology is fully general and can be applied to any dataset. Our procedure enables the standardization and optimization of the measurement of stellar velocity dispersions. This is a crucial step required for using them for cosmography, and it will be adopted as standard by the TDCOSMO collaboration.

Looking ahead, it is important to note that when using a measured $\sigma$ to estimate velocity moments for dynamical models, there are additional sources of potential systematics and model-dependent effects that must be evaluated, mitigated, or corrected, to ensure that a $1\%$ accuracy on $\sigma$ translates to our target of $2\%$ on H$_0$. Such considerations include how accurately a Gaussian line-of-sight velocity distribution (LOSVD) approximates the true velocity moment and the corresponding accuracy and precision in mass estimates derived from such models, as well as triaxiality and projection effects \citet[][]{tdcosmo_xxi_triaxiality}. These aspects will be addressed in future papers, building upon the key result of this paper, namely, the ability to accurately and precisely determine the observable quantity, $\sigma$.

\section{Data availability}

To facilitate the adoption of our method by third parties, the list of templates and the scripts used in this analysis has been made publicly available through the GitHub repository available at URL \url{https://github.com/TDCOSMO/KINEMATICS_METHODS}. We have also developed the \textsc{squirrel} pipeline to implement our method. The pipeline was first introduced in \citet{Shajib25} and has been made publicly available at \url{https://github.com/ajshajib/squirrel}. \textsc{squirrel} is built on the penalized PiXel Fitting (\textsc{pPXF}) software package \citep{Cappellari17, Cappellari23}.

\begin{acknowledgements}
The first two authors should be regarded as joint first authors.
We are grateful to S.~Trager and A.~Vazdekis for sharing with us their quality flags of the XSL and MILES libraries, respectively.
We thank A.S.~Bolton and M.Bernardi for useful discussions regarding SDSS velocity dispersions. 
We thank E.~Buckley-Geer, M.~Millon, and all the friends of the TDCOSMO collaboration for useful feedback that improved this manuscript.
This work is based on observations made with the NASA/ESA/CSA James Webb Space Telescope. The data were obtained from the Mikulski Archive for Space Telescopes at the Space Telescope Science Institute, which is operated by the Association of Universities for Research in Astronomy, Inc., under NASA contract NAS 5-03127 for JWST. These observations are associated with program \#1974.
Some of the data presented herein were obtained at Keck Observatory, which is a private 501(c)3 non-profit organization operated as a scientific partnership among the California Institute of Technology, the University of California, and the National Aeronautics and Space Administration. The Observatory was made possible by the generous financial support of the W.~M.~Keck Foundation. 
The authors wish to recognize and acknowledge the very significant cultural role and reverence that the summit of Maunakea has always had within the Native Hawaiian community. We are most fortunate to have the opportunity to conduct observations from this mountain.
Based in part on observations collected at the European Southern Observatory data obtained from the ESO Science Archive Facility.
Some of the data presented here were taken from the Sloan Digital Sky Survey Archive.
Support for this work was provided by NASA through the NASA Hubble Fellowship grant HST-HF2-51492 awarded to AJS by the Space Telescope Science Institute, which is operated by the Association of Universities for Research in Astronomy, Inc., for NASA, under contract NAS5-26555. AJS also received support from NASA through the STScI grants HST-GO-16773 and JWST-GO-2974.
We acknowledge support by NSF through grants NSF-AST-1906976, NSF-AST-2407277, and NSF-AST-1836016, and from the Moore Foundation through grant 8548.
We acknowledge support from NASA through grants JWST-GO-1974 and JWST-GO-2974.
SB thanks the Department of Physics and Astronomy, Stony Brook, for their support.
\\
\end{acknowledgements}

\bibliographystyle{aa} 
\bibliography{main}

\begin{appendix}

\section{Comparisons of template libraries for various samples}

\begin{figure*}
    \includegraphics[width=\textwidth]{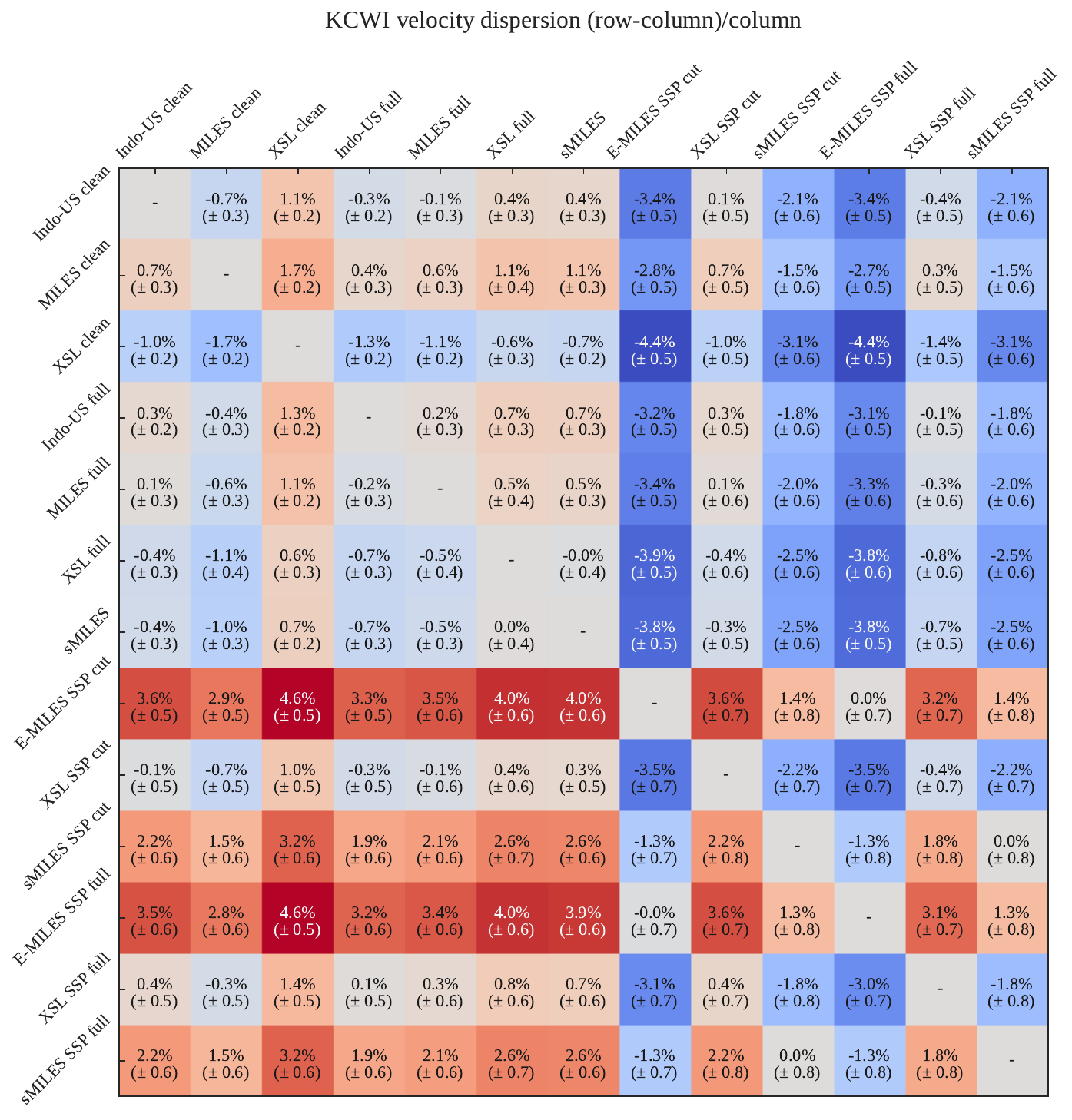}
    \caption{Comparison of velocity dispersion measurements of the KCWI sample using a range of template libraries. The wavelength range is 3600--4500 \AA. This figure is in the same format as Figure~\ref{fig:comparison_all_stellar_ssp}.}
    \label{fig:kcwi_library_comparison}
\end{figure*}

\begin{figure*}
    \includegraphics[width=\textwidth]{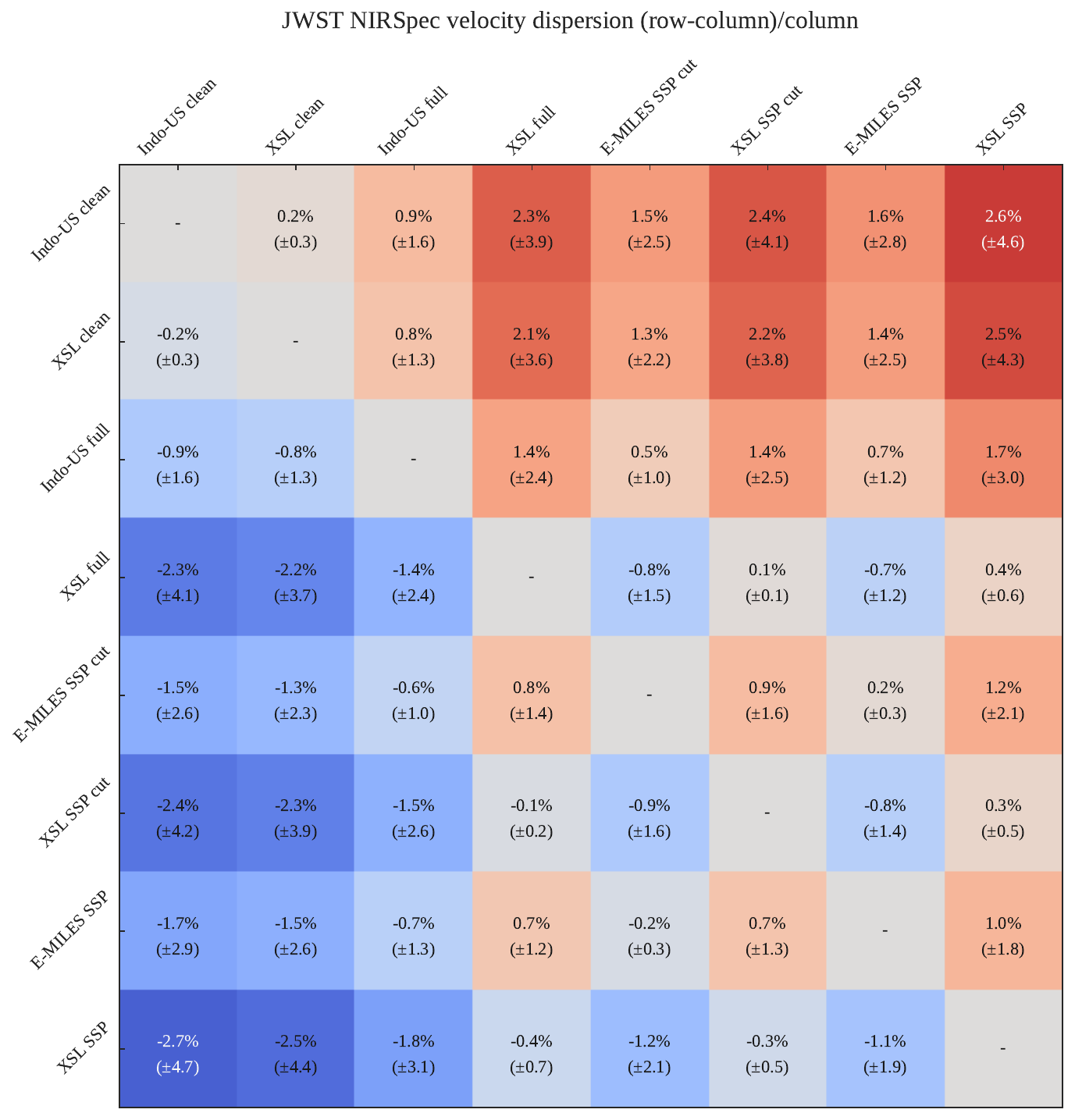}
    \caption{\label{fig:nirspec_comparison}
    Comparison of velocity dispersion measurements for the JWST/NIRSpec spectra for \lensname. The differences in the measured dispersions are averaged over 21 spaxels within a 0\farcs5$\times$0\farcs5 square at the center of the lens galaxy. This figure is in the same format as Figure~\ref{fig:comparison_all_stellar_ssp}, except that fewer libraries are shown, only the ones that include the wavelength region of the Ca \textsc{ii} triplet used to fit the JWST/NIRSpec data.}
\end{figure*}

Here, we show the comparison of velocity dispersions across template libraries as in Figure~\ref{fig:comparison_all_stellar_ssp} for the KCWI, SDSS, and JWST/NIRSpec samples in Figures~\ref{fig:kcwi_library_comparison} to~\ref{fig:nirspec_comparison}.

\begin{figure*}
    \includegraphics[width=\textwidth]{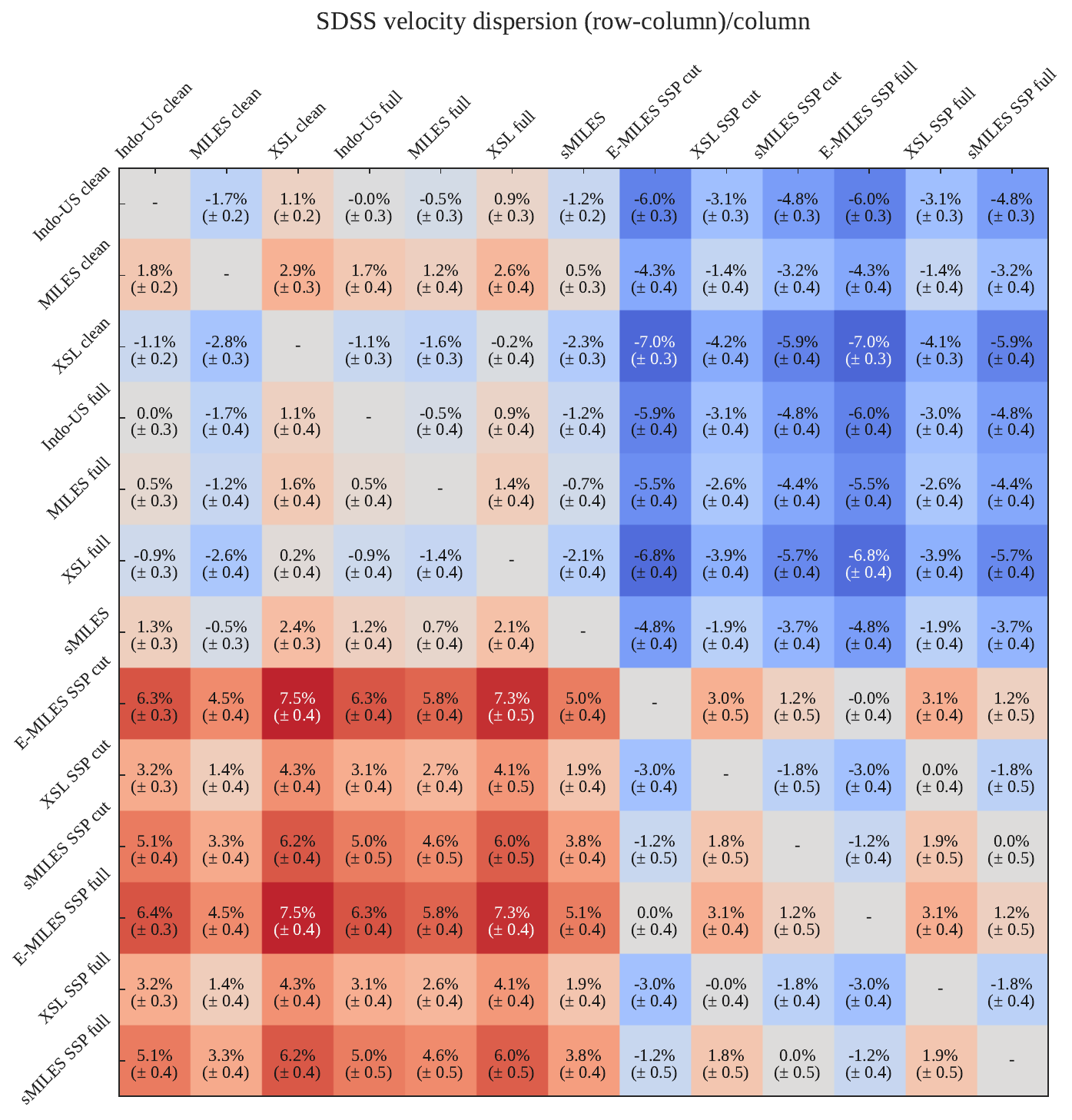}
    \caption{Comparison of velocity dispersion measurements of the sample with SDSS spectra with S/N$>$15 \AA$^{-1}$, using a range of template libraries. The wavelength range is 3600-5350 \AA. This figure is in the same format as Figure~\ref{fig:comparison_all_stellar_ssp}.}
\end{figure*}

\section{Validation on mock data}
\label{app:mock_data}

We validated the accuracy of the methods described in Section~\ref{sec:recipe} by applying the recipe to a mock dataset. We generated two samples of mock spectra. In the first, mock spectra are based on stellar templates from XSL-DR3 and designed to simulate spectra from a KCWI observation. The resolution of XSL templates (13 \kmps) is better than the target resolution of KCWI data (35 \kmps), so the effect of the instrumental line spread function is simulated. We simulated KCWI data because the sample's high S/N (mean 160 \AA$^{-1}$) allows us to probe the accuracy of our methods over a range of realistic noise levels. The second test is identical to the first except for the following: the mock spectra are based on templates from the Indo-US library at the native template resolution of $\sim43$ \kmps.

To begin, we selected a representative object from our KCWI galaxy sample, SDSSJ1204+0358. This object has a typical velocity dispersion for lens galaxies ($\sigma_{\rm KCWI}\sim250$ \kmps). From the fiducial fits to this object using both the XSL and Indo-US libraries, in the wavelength range 3600-4500$\AA$ with additive and multiplicative polynomials of 6 and 2, we collected the weights of the templates used for the fit. The XSL fit utilized 13 of the available 496 templates with normalized weights greater than 1\%, while the Indo-US fit utilized 16 of the available 989 templates. These templates were log-sampled in wavelength. For the mock XSL sample, templates are convolved to the resolution of KCWI. From the subset of 13 (16) templates that compose the original combination in the XSL (Indo-US) fit, we constructed three new mock spectra. We randomly alter the weights of each template by 10\% and take a spectrum that results in a root-mean-square residual (averaged across wavelengths) < 10\% compared with the original combination, to avoid unrealistic alterations to the stellar population represented by the linear combination of templates. We oversampled the spectrum by a factor of 12 to ensure an accurate convolution with the Gaussian LOSVD, which has a mean velocity of zero and a velocity dispersion of 300 \kmps. The resulting convolved spectrum is rebinned back to KCWI resolution. For each of the three mock spectra, we generated 100 noise realizations for each of seven different S/N per $\AA$: 
15, 30, 45, 60, 90, 120, and 150, by dividing the flux of the spectra by the desired S/N, and by the square root of the KCWI pixel size ($0.5\AA/ \rm pixel$) for the XSL sample. We tested the recipe outlined in Section~\ref{sec:recipe} on these mock samples, each consisting of 2100 spectra. Our goal is to assess the method's ability to constrain the systematic uncertainties introduced by the template libraries on three bases, determining whether:

\begin{enumerate}
    \item  Our method can correctly identify the most appropriate source of templates, namely, the parent template library used to generate the mock sample;
    \item  The resulting weighted averages result in the correct velocity dispersion within $1\%$;
    \item The S/N threshold of 30 per $\AA$
    results in systematic errors within 1$\%$, as we have found for our observed datasets.
\end{enumerate}

For each mock spectrum, we fit in the wavelength range 3600-4500$\AA$ with the clean Indo-US, MILES, and XSL libraries. We did not introduce polynomial corrections to the continuum, since they were not included in the mock galactic spectra.  Figure~\ref{fig:stellar-spectra} shows that there is a mismatch in continuum shape between the template libraries for the same observed stars. For the mock XSL sample, the correction is not needed for fits that utilize XSL templates, and a high polynomial order could bias the resulting velocity dispersion while at the same time not being high enough to properly account for the Indo-US and MILES fits. The resolutions of Indo-US and MILES are lower than the mock XSL spectra, so the templates cannot be broadened to the data resolution. Instead, the correction is applied in quadrature after the measurement utilizing Equation~\ref{eq:sigma_quad}. For the mock Indo-US spectra, the same post-measurement correction is applied only for MILES. For each sample and each S/N, we combine the subsample of 300 mock spectra with and without the BIC-weighting methods described in Section~\ref{sec:recipe}. The Bessel correction is implemented for both weighted and unweighted systematic uncertainties. 

Weights were calculated from the $\rm \Delta BIC$ across each S/N subsample and are given for the XSL and Indo-US mock samples in Tables~\ref{table:mock_xsl_weights} and~\ref{table:mock_indo_weights}, respectively. For the XSL mock sample, on average across the S/N subsamples, the BIC-weighting prefers XSL with 99.97$\%$ of the weight, with the other 0.03$\%$ going to Indo-US. This strong preference is expected. On the other hand, the Indo-US mock sample only mildly prefers its parent template library, with almost equal weight given to MILES and about half that weight given to XSL for all subsamples with S/N per $\AA>30$. This might be because the mock spectra are sampled at lower resolution than the XSL mock sample.

\begin{figure*}
\includegraphics[width=\textwidth]{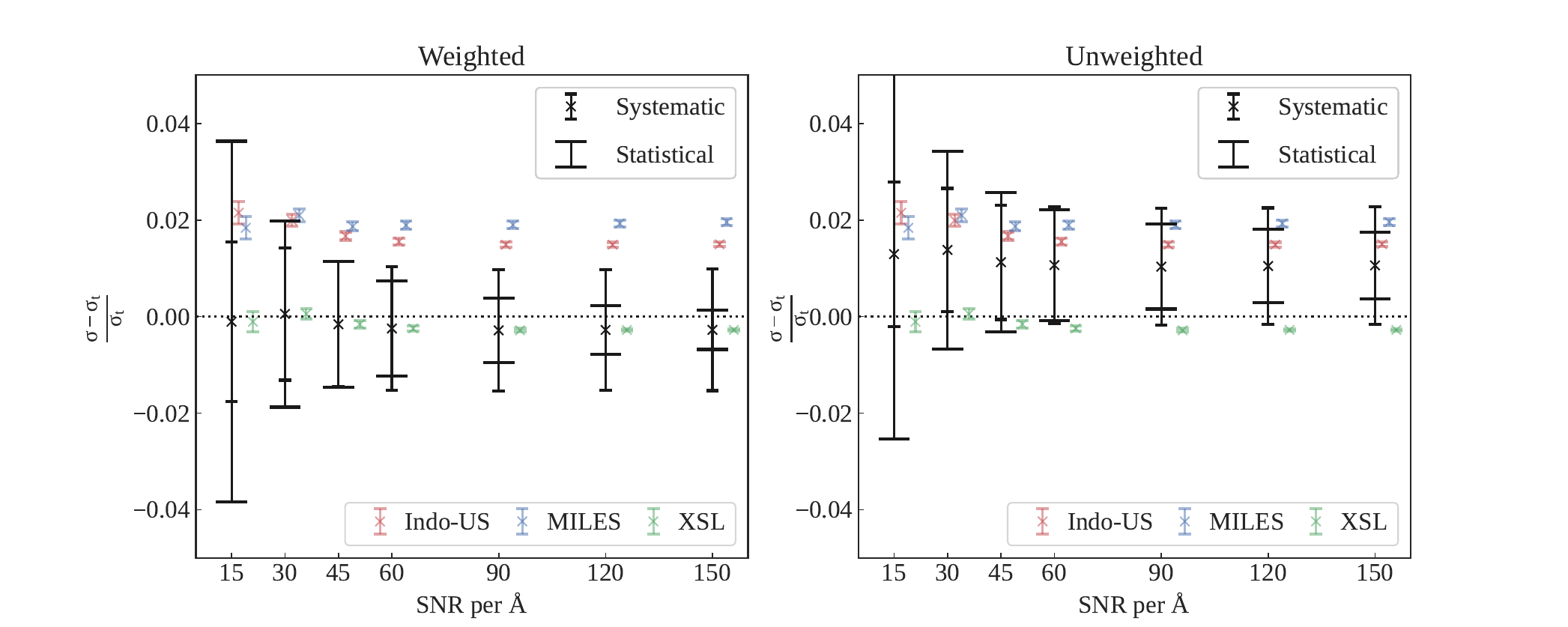}
        \caption{\label{fig:mock_xsl}
    Results on mock data generated from XSL.}
\end{figure*}

\begin{figure*}
\includegraphics[width=\textwidth]{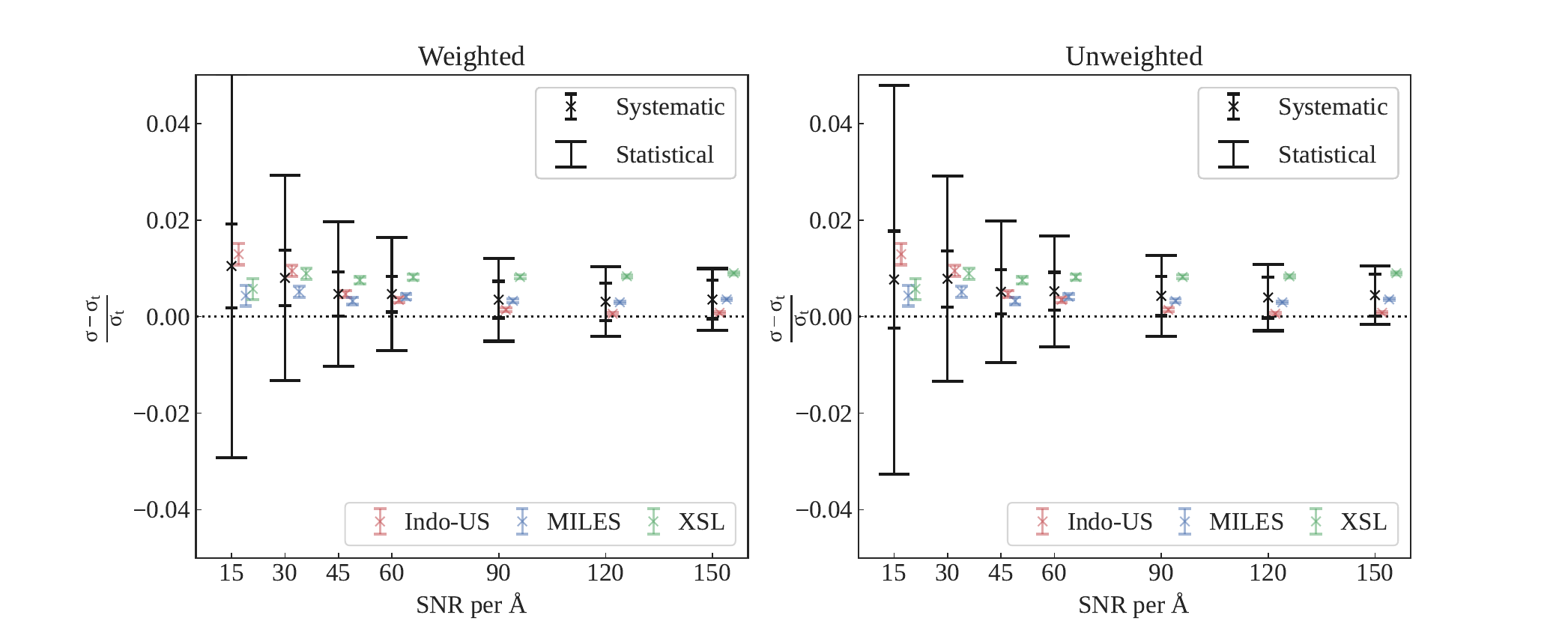}
        \caption{\label{fig:mock_indo}
    Results on mock data generated from Indo-US.}
\end{figure*}

Results for the XSL and Indo-US mock samples are listed in Tables~\ref{table:mock_xsl} and~\ref{table:mock_indo} shown in Figures~\ref{fig:mock_xsl} and~\ref{fig:mock_indo}, where the left panel shows the weighted results and the right panel shows the unweighted results. In both panels, the x-axis shows the S/N per $\AA$,
and the y-axis shows the fractional deviation from the true value of 300 \kmps. The red, blue, green markers show the unweighted mean values for Indo-US, MILES, and XSL fits, respectively, for the given S/N subsample. Black markers show the mean value over the template libraries combined over the whole subsample, with two error bars each, representing the systematic errors with narrow horizontal cap-width and the statistical errors with the comparatively wider horizontal cap-width.

For both mock samples, the weighted mean velocity dispersions are within $\sim1\%$ of the mean for all S/N subsamples, indicating that the method does consistently return sub-percent accuracy. For the XSL mock sample, the unweighted mean velocity dispersions are unable to improve accuracy to within 1\%, regardless of S/N. Fits to the XSL mock sample with Indo-US and MILES libraries overestimate the true velocity dispersion by $\sim2\%$, which is likely because they might require some polynomial corrections. On the other hand, the Indo-US mock sample is accurate to within 1\% for the unweighted case as well.

In terms of estimating the systematic errors, the XSL mock sample results in slightly overestimated systematic uncertainties for the weighted case in comparison with the unweighted case. This is a result of the extreme imbalance in the weights, which inflates the uncertainty through the Bessel correction (see Equation~\ref{eq:sys_error_Bessel}), where the term explodes when the sum of weights approaches 1. This is highly unlikely to occur for real data, where our results clearly show sub-percent systematic uncertainties (see Section~\ref{sec:results}). Even in this odd scenario, the unweighted and weighted estimate of the systematic errors reach 1.2 and 1.25\%, and the BIC-weighting is important for ensuring sub-percent accuracy. The Indo-US mock sample results in sub-percent estimates of the systematic uncertainties for all S/N. Across the subsamples, there is a slight improvement in the estimate over the unweighted case.

In summary, our tests with mock spectra indicate the following:
\begin{enumerate}
    \item Our method returns the true velocity dispersion to within 1\% accuracy for all S/N subsamples when utilizing the BIC weights and approaches this threshold even without using the weights;
    \item Our method correctly identifies the most appropriate set of stellar templates, the parent library from which the mock spectra are generated;
    \item For S/N per $\AA=30$ and above the results are accurate within 1\%, as we find in real data. The weighted estimate of the systematic error is within 1\% for the Indo-US mock, while it is slightly above 1\% for the XLS mock, owing to numerical issues with the Bessell correction when the weight of the preferred library is so close to unity.

\end{enumerate}

\begin{table*}
\caption{Library weights for XSL mock template fits for each S/N subsample.}
\begin{tabular}{l c c c}
\hline
S/N per $\AA$ & Indo-US & MILES & \textbf{XSL} \\
\hline
15 & 0.0006 & 0.0001 & 0.9994 \\
30 & 0.0000 & 0.0000 & 1.0000 \\
45 & 0.0000 & 0.0000 & 1.0000 \\
60 & 0.0000 & 0.0000 & 1.0000 \\
90 & 0.0000 & 0.0000 & 1.0000 \\
120 & 0.0003 & 0.0000 & 0.9997 \\
150 & 0.0015 & 0.0000 & 0.9985 \\
\end{tabular}
\label{table:mock_xsl_weights}
\end{table*}

\begin{table*}
\caption{Library weights for Indo-US mock template fits for each S/N subsample.}
\begin{tabular}{l c c c}
\hline
S/N per $\AA$ & \textbf{Indo-US} & MILES & XSL \\
\hline
15 & 0.6587 & 0.0057 & 0.3355 \\
30 & 0.4635 & 0.3048 & 0.2317 \\
45 & 0.4065 & 0.3902 & 0.2033 \\
60 & 0.4038 & 0.3944 & 0.2019 \\
90 & 0.4016 & 0.3976 & 0.2008 \\
120 & 0.4012 & 0.3981 & 0.2006 \\
150 & 0.4012 & 0.3982 & 0.2006 \\
\end{tabular}
\label{table:mock_indo_weights}
\end{table*}

\begin{table*}
\caption{Results from mock data generated from XSL for each S/N subsample.}
\resizebox{\textwidth}{!}{
\begin{tabular}{l c c c | c c c } 
    \hline
S/N per $\AA$ & $<\bar \sigma_w-\sigma_t>/\sigma_t$ & $<\Delta_B\sigma_w/\bar \sigma_w>$ & $<\delta\sigma_w/\bar \sigma_w>$ & $<\bar \sigma_u-\sigma_t>/\sigma_t$ & $<\Delta_B\sigma_u/\bar \sigma_u>$ & $<\delta\sigma_u/\bar \sigma_u>$  \\
\hline
15 & -0.11\% & 1.65\% & 3.73\% & 1.29\% & 1.52\% & 3.88\% \\
30 & 0.05\% & 1.37\% & 1.93\% & 1.38\% & 1.30\% & 2.08\% \\
45 & -0.16\% & 1.29\% & 1.30\% & 1.13\% & 1.20\% & 1.46\% \\
60 & -0.25\% & 1.27\% & 0.98\% & 1.07\% & 1.22\% & 1.16\% \\
90 & -0.28\% & 1.25\% & 0.67\% & 1.03\% & 1.22\% & 0.89\% \\
120 & -0.28\% & 1.25\% & 0.51\% & 1.05\% & 1.22\% & 0.77\% \\
150 & -0.27\% & 1.26\% & 0.40\% & 1.06\% & 1.23\% & 0.70\% \\
\bottomrule
\end{tabular}
}
\label{table:mock_xsl}
\tablefoot{
Columns 2-4 show values utilizing the weights given in Table~\ref{table:mock_xsl_weights}. Column 2 gives the accuracy of the mean velocity dispersions with respect to the true velocity dispersion $\sigma_t = 300$ \kmps. Column 3 gives the Bessel-corrected average systematic uncertainties. Column 4 gives the average statistical uncertainties. Columns 5-7 show the same values calculated with equal weights.
}
\end{table*}

\begin{table*}
\caption{Results from mock data generated from Indo-US for each S/N subsample. }
\resizebox{\textwidth}{!}{
\begin{tabular}{l c c c | c c c } 
\hline
S/N per $\AA$ & $<\bar \sigma_w-\sigma_t>/\sigma_t$ & $<\Delta_B\sigma_w/\bar \sigma_w>$ & $<\delta\sigma_w/\bar \sigma_w>$ & $<\bar \sigma_u-\sigma_t>/\sigma_t$ & $<\Delta_B\sigma_u/\bar \sigma_u>$ & $<\delta\sigma_u/\bar \sigma_u>$  \\
\hline
15 & 1.05\% & 0.88\% & 4.01\% & 0.77\% & 1.01\% & 4.06\% \\
30 & 0.80\% & 0.58\% & 2.15\% & 0.78\% & 0.59\% & 2.14\% \\
45 & 0.47\% & 0.45\% & 1.50\% & 0.51\% & 0.47\% & 1.48\% \\
60 & 0.47\% & 0.37\% & 1.18\% & 0.52\% & 0.39\% & 1.16\% \\
90 & 0.35\% & 0.38\% & 0.86\% & 0.43\% & 0.41\% & 0.84\% \\
120 & 0.31\% & 0.39\% & 0.72\% & 0.40\% & 0.43\% & 0.69\% \\
150 & 0.35\% & 0.40\% & 0.65\% & 0.44\% & 0.44\% & 0.61\% \\
\bottomrule
\end{tabular}
}
\label{table:mock_indo}
\tablefoot{
Columns 2-4 show values utilizing the weights given in Table~\ref{table:mock_indo_weights}. Column 2 gives the accuracy of the mean velocity dispersions with respect to the true velocity dispersion $\sigma_t = 300$ \kmps. Column 3 gives the Bessel-corrected average systematic uncertainties. Column 4 gives the average statistical uncertainties. Columns 5-7 show the same values calculated with equal weights.
}
\end{table*}

\section{Derivation of weighted combination formulae}
\label{app:Bessel}

\newcommand{\expct}[1]{\mathbb{E}\left[{{#1}}\right]}
\newcommand{\var}[1]{\text{Var}\left({{#1}}\right)}
\newcommand{\terma}[1]{\textcolor{BurntOrange}{{#1}}}
\newcommand{\termb}[1]{\textcolor{DarkOrchid}{{#1}}}
\newcommand{\termc}[1]{\textcolor{ForestGreen}{{#1}}}

Here, we provide a derivation of Equations \ref{eq:mean}, \ref{eq:stat_error}, \ref{eq:sys_error_Bessel}, and \ref{eq:covariance_Bessel} using variance algebra. We can formulate the problem to solve by first taking a number of random variables $X_k$ with associated weights $w_k$. Let $Y$ be another random variable that is sampled from $X_k$ with probability $p_k = w_k / \sum_k w_k$. For our use case, $X_k$ is the velocity dispersion distribution from a given template library, and $Y$ is the combined velocity dispersion distribution with weights $w_k$. Then, the problem is to derive $\expct{Y}$ and $\var{Y}$.

For simplicity, let us denote $\expct{X_k} \coloneqq \mu_k$ and $\var{X_k} \coloneqq \sigma^2_k$. Then, we have $\expct{Y \mid K=k} = \mu_k$ and $\var{Y \mid K=k} = \sigma^2_k$.
From the law of total expectation, we obtain
\begin{equation}
    \expct{Y} = \expct{\expct{Y \mid K}} = \sum_k p_k \mu_k,
\end{equation}
which leads to Equation \ref{eq:mean}. Similarly, from the law of total variance, we obtain
\begin{equation}
    \var{Y} = \expct{\var{Y \mid K}} + \var{\expct{Y \mid K}},
\end{equation}
Here, the first term on the right-hand side above is the unexplained variance, which we call the statistical uncertainty, and the second term is the explained variance, which we refer to as the systematic uncertainty. The unexplained variance can be expressed as
\begin{equation}
    \expct{\var{Y \mid K}} = \sum_k p_k \sigma^2_k,
\end{equation}
which leads to Equation \ref{eq:stat_error}.

To obtain the expression for the explained variance $\var{\expct{Y \mid K}} \coloneqq \var{\mu_k}$, let us first define
\begin{equation}
    s^2 \equiv \sum_k p_k \left(\mu_k - \sum_j p_j \mu_j \right)^2.
\end{equation}
Then, the expected value of $s^2$ is
\begin{equation}
    \begin{aligned}
    \expct{s^2} &= \expct{\sum_k p_k \left( \terma{\mu_k^2} + \termb{\sum_{i,j} p_i p_j\, \mu_i \mu_j} - \termc{2 \sum_j p_j\, \mu_j \mu_k} \right)} \\
    &= \sum_k p_k \left(\terma{\expct{\mu_k^2}} + \termb{\sum_{i, j} p_i p_j \, \expct{\mu_i \mu_j} } - \termc{2 \sum_j p_j \expct{\mu_j \mu_k}} \right) \\
    &= \sum_k p_k \left\{ \terma{\var{\mu_k} + \expct{\mu_k}^2} \phantom{\sum_j} \right. \\
    &\qquad\qquad + \termb{\sum_{i, j} p_i p_j \left( \var{\mu_i}\delta_{i,j} + \expct{\mu_i} \expct{\mu_j} \right)} \\
    &\qquad\qquad - \left. \termc{2 \sum_{j} p_j \left( \var{\mu_k}\delta_{j, k} + \expct{\mu_j} \expct{\mu_k} \right)} \right\} \\
    &=\sum_k p_k \left\{\terma{\var{\mu_k} + \cancel{\expct{\mu_k}^2}} \phantom{\sum_j} \right. \\
    &\qquad\qquad + \termb{\var{\mu_k} \sum_j p_j^2 + \cancel{\expct{\mu_k}^2 \sum_{i, j} p_i p_j}} \\
    &\qquad\qquad \left. - \termc{ 2p_k\, \var{\mu_k} - \cancel{2 \expct{\mu_k}^2 \sum_j p_j}} \right\} \\
    &= \var{\mu_k}  \sum_k p_k \left(\terma{1} + \termb{\sum_j p_j^2} - \termc{2p_k} \right) = \var{\mu_k} \left(1 - \sum_k p_k^2 \right).
    \end{aligned}
\end{equation}
Here, we have used the relation $\sum_k p_k = 1$, which is true by definition. Therefore, an unbiased estimate (i.e., with the Bessel correction applied) of the explained variance is obtained by
\begin{equation}
    \var{\expct{Y \mid K}} = \frac{s^2}{1 - \sum_k p_k^2} = \frac{\sum_k w_k \left(\mu_k - \expct{\mu_k} \right)^2}{\sum_k w_k - \sum_k w_k^2 / \sum_k w_k },
\end{equation}
which leads to Equation \ref{eq:sys_error_Bessel}. Equation \ref{eq:covariance_Bessel} can also be derived following the same procedure.

\end{appendix}

\end{document}